%% file: paper.tex
\documentclass[pdflatex,sn-basic]{sn-jnl}% Basic Springer Nature Reference Style/Chemistry Reference Style
\usepackage{color,bm,natbib}

\input macros
\input journals

\graphicspath{{./fig/}{./png/}}

\jyear{2022}%

\begin{document}

\title[Simulations of solar and stellar dynamos and their theoretical interpretation]{Simulations of solar and stellar dynamos and their theoretical interpretation}

\author*[1,2]{\fnm{Petri J.} \sur{K\"apyl\"a}}\email{pkapyla@leibniz-kis.de}

\author[3]{\fnm{Matthew K.} \sur{Browning}}\email{M.K.M.Browning@exeter.ac.uk}

\author[4]{\fnm{Allan Sacha} \sur{Brun}}\email{sacha.brun@cea.fr}

\author[5,6]{\fnm{Gustavo} \sur{Guerrero}}\email{guerrero@fisica.ufmg.br}

\author[7]{\fnm{J\"orn} \sur{Warnecke}}\email{warnecke@mps.mpg.de}

\affil*[1]{\orgname{Leibniz Institute for Solar Physics (KIS)}, \orgaddress{\street{Sch\"oneckstra{\ss}e 6}, \city{Freiburg}, \postcode{79104}, \country{Germany}}}

\affil[2]{\orgdiv{Institute for Astrophysics and Geophysics}, \orgname{University of G\"ottingen}, \orgaddress{\street{Friedrich-Hund-Platz 1}, \city{G\"ottingen}, \postcode{37077}, \country{Germany}}}

\affil[3]{\orgdiv{Department of Physics \& Astronomy}, \orgname{University of Exeter}, \orgaddress{\street{Stocker Road}, \city{Exeter}, \postcode{EX4 4QL}, \country{United Kingdom}}}

\affil[4]{\orgdiv{D\'epartement d’Astrophysique/AIM, CEA/IRFU, CNRS/INSU}, \orgname{Univ. Paris-Saclay \& Univ. de Paris Cit\'e}, \orgaddress{\city{Gif-sur-Yvette}, \postcode{91191}, \country{France}}}

\affil[5]{\orgdiv{Physics Department}, \orgname{Universidade Federal de Minas Gerais}, \orgaddress{\street{Av. Antonio Carlos 6627}, \city{Belo Horizonte}, \postcode{MG 31270-901}, \country{Brazil}}}

\affil[6]{\orgdiv{Physics Department}, \orgname{New Jersey Institute of Technology}, \orgaddress{\street{323 Dr Martin Luther King Jr Blvd}, \city{Newark}, \postcode{NJ 07103}, \country{USA}}}

\affil[7]{\orgname{Max Planck Institute for Solar System Research}, \orgaddress{\street{Justus-von-Liebig-Weg 3}, \city{G\"ottingen}, \postcode{37077}, \country{Germany}}}


\abstract{We review the state of the art of three dimensional
  numerical simulations of solar and stellar dynamos. We summarize
  fundamental constraints of numerical modelling and the techniques to
  alleviate these restrictions. Brief summary of the relevant
  observations that the simulations seek to capture is given. We
  survey the current progress of simulations of solar convection and
  the resulting large-scale dynamo. We continue to studies that model
  the Sun at different ages and to studies of stars of different
  masses and evolutionary stages. Both simulations and observations
  indicate that rotation, measured by the Rossby number which is the
  ratio of rotation period and convective turnover time, is a key
  ingredient in setting the overall level and characteristics of
  magnetic activity. Finally, efforts to understand global 3D
  simulations in terms of mean-field dynamo theory are discussed.}


\maketitle

\section{Introduction}\label{sec1}

The intriguing coherence of the solar magnetic cycle has fascinated
researchers for more than a century starting from Hale's discovery of
magnetic field in sunspots
\citep{1908ApJ....28..315H,1919ApJ....49..153H}, and early attempts to
build simple models \citep{Larmor1919,1933MNRAS..94...39C}. The first
successful models of the solar cycle made use of mean-field
approximations yielding equations where only the large-scale
contributions were explicitly computed, whereas the small scales were
characterised by physically plausible parameterizations
\citep{Pa55b,1969AN....291...49S}. Mean-field models opened up new
avenues in studying solar and stellar magnetism but their Achilles'
heel is the parameterizations of the small scales which are in general
untractable analytically in parameter regimes relevant to stars. This
is due to the closure problem of turbulence rendering such models
susceptible to fine-tuning.  A review of modern mean-field theory is
presented elsewhere in this volume \citep{2023arXiv230312425B}.

Rapidly increasing computing power allowed for the first direct
solutions of the equations of (magneto)hydrodynamics in spherical
shells in the late 1970s and early 1980s
\citep{Gi77,1981ApJS...46..211G,Gi83,Gl85}. Prior to the these
simulations and the discovery of the internal rotation profile of the
Sun, the angular velocity was generally assumed to decrease as a
function of radius, in which case the propagation of the dynamo wave
from a mean-field $\alpha\Omega$ dynamo is predicted to be equatorward
given typical assumptions regarding the influence of the Coriolis
force on convective eddies \citep{Pa55b,1969AN....291...49S,Yo75};
see, however, \cite{RS72}. This changed definitively when
helioseismology revealed that the angular velocity is actually
increasing with radius in the bulk of the solar convection zone
\citep[e.g.][]{1984Natur.310...22D,Schouea98} which lead to the
``dynamo dilemma'' \citep{Pa87a}. This dilemma was also captured by
the early 3D simulations where solar-like differential rotation with
fast equator and slow poles was qualitatively reproduced, but where
the dynamo waves propagated toward the poles, contrary to the Sun
\citep{Gi83,Gl85}.

After this, the interest in 3D simulations of solar and stellar
dynamos waned and was not rekindled until the early 2000s, starting
with the development of the ASH (Anelastic Spherical Harmonic) code
\citep[e.g.][]{METCGG00,2000ApJ...533..546E,2002ApJ...570..865B}. While
many of the early studies concentrated on the Sun
\citep[e.g.][]{BMT04,BMBT06,2008ApJ...673..557M}, a proliferation of
models from various groups using different codes occurred in the 2010s
when simulations of more rapidly rotating Suns started to yield cycles
and equatorward migration more or less routinely
\citep[e.g.][]{GCS10,KKBMT10,BMBBT11,KMB12,NBBMT13,ABMT15,MMK15,2015ApJ...810...80S}.
Furthermore, simulations of main-sequence stars other than the Sun
also started to appear covering the mass range from fully convective M
dwarfs
\citep[e.g.][]{DSB06,2008ApJ...676.1262B,YCMGRPW15,2020ApJ...893..107B,2021A&A...651A..66K}
to F stars with thin surface convection zones
\citep{2013ApJ...777..153A, 2022A&A...667A..43B}, as well as core
convection, dynamos, and interaction with fossil fields in more
massive A, B and O stars
\citep[e.g.][]{2009ApJ...705.1000F,2016ApJ...829...92A}. Models
exploring stellar magnetism outside of the main sequence have also
started to appear, including pre-main sequence stars
\citep[e.g.][]{Emeriau+17}, red giants
\citep[e.g.][]{2004A&A...423.1101D,BP09}, and newly born neutron stars
\citep[e.g.][]{2020SciA....6.2732R,2022ApJ...924...75M}.

Parallel to the developments in simulations, observational data and
knowledge regarding stellar magnetism has also experienced explosive
growth. We now have dozens of stars with observed cycles from
long-term observing campaigns monitoring chromospheric emission
\citep[e.g.][]{1995ApJ...438..269B}. However, the systematics of these
cycles as a function of stellar rotation are still under debate
\citep[e.g.][]{2017ApJ...845...79B,2018A&A...616A.108B,2018A&A...619A...6O,Bonanno+22}.
Zeeman-Doppler imaging has also revealed polarity reversals
\citep[e.g.][]{KMHI13,2018A&A...620L..11B}, as well as large-scale
non-axisymmetric and dipole-dominated magnetic fields in rapidly
rotating late-type stars \citep[e.g.][]{2021A&ARv..29....1K}. Finally,
magnetic activity saturates when the stellar Rossby number $\Ro_\star
= \Prot/\tauconv$, which is the ratio of the rotation period and the
convective turnover time, is less than about 0.1, such that for lower
$\Ro$ the activity and magnetic field strength is roughly constant
\citep[e.g.][]{2018MNRAS.479.2351W,2022A&A...662A..41R}. These basic
observations are crucial constraints for the numerical
simulations. Nevertheless, the Sun still poses the stringest
constraints to simulations due its proximity and access to its
interior structure through helioseismology. Somewhat surprisingly the
current 3D simulations struggle to reproduce not only the dynamo, but
also the convective amplitudes and the differential rotation of the
Sun, often yielding anti-solar (slow equator, fast poles) solutions
with nominally solar luminosity and rotation rate
\citep[e.g.][]{2011AN....332..897M,KKB14,GYMRW14,HRY15a,2017ApJ...836..192B}. This
issue has been dubbed the convective conundrum
\citep{2016AdSpR..58.1475O} and poses arguably the greatest challenge
in the field of stellar dynamo simulations today. It has also been
suggested that that the Sun is close to a transition where its dynamo
efficiency diminishes \citep[e.g.][]{2016Natur.529..181V}, possibly
due to a shift from solar-like to anti-solar differential rotation,
making it difficult to capture by simulations
\citep[e.g.][]{KKB14,2022ApJ...926...21B}. Our aim in the following is
to review the current successes and shortcomings of current
simulations in capturing the relevant observations.

The remainder of the review is organised as follows: the basic
equations and physics are discussed in \Sec{sec:physics} and the
limitations of the numerical approach are reviewed in
\Sec{sec:numapp}. The relevant observations and the main results of
current 3D simulations of various types of stars are reviewed in
\Sec{sec:observations} and \Sec{sec:results}, respectively.
\Sec{sec:conMF} gives an overview of the comparisons between
mean-field models and global simulations. Finally, we conclude in
\Sec{sec:conclusions} with an overview of the state of the field,
current challenges, and possible future directions.

\section{Relevant physics and equations}\label{sec:physics}

Stellar convection zones are described by the equations of
magnetohydrodynamics (MHD), describing the time evolution of the
magnetic field and conservation of mass, momentum, and energy:
\begin{eqnarray}
  \frac{\pd \BBB}{\pd t} &=& \bm\nabla\times(\uuu \times \BBB - \eta \mu_0 \JJJ), \\
  \frac{\pd \rho}{\pd t} &=& - \bm\nabla\bm\cdot(\rho\uuu),\\
  \rho \frac{\pd \uuu}{\pd t} &=& - \bm\nabla \bm\cdot (\rho \uuu \uuu) + \rho \gggg - \bm\nabla p - 2 \rho \bm\Omega_0 \times \UUU + \JJJ \times \BBB + \bm\nabla\bm\cdot \FFF^{\rm visc}, \\
  \rho T \frac{\pd s}{\pd t} &=& - \bm\nabla \bm\cdot (\rho s \uuu) + \bm\nabla\bm\cdot \bm{\mathcal{F}} + {\cal H},
\end{eqnarray}
where $\BBB$ is the magnetic field, $\uuu$ the velocity, $\eta$ is the
magnetic diffusivity, $\mu_0$ the permeability of vacuum, $\JJJ =
\mu_0^{-1} \bm\nabla \times \BBB$ is the current density, $\rho$ is
the fluid density, $\gggg=-\bm\nabla\phi$ is the acceleration due to
gravity, where $\phi$ is the gravitational potential, $p$ is the gas
pressure, $\bm\Omega_0$ is the rotation rate of the star, $\FFF^{\rm
  visc}$ is the viscous force, $s$ is the specific entropy, and
$\bm{\mathcal{F}} = \bm{\mathcal{F}}^{\rm rad} + \bm{\mathcal{F}}^{\rm
  SGS}$ describes radiative and any subgrid-scale (SGS) fluxes that
are present. ${\cal H}$ describes additional cooling and heating that
is sometimes used instead of, or in addition to, the radiative flux
\citep[e.g.][]{GCS10,2019ApJ...880....6G,2019ApJ...871..217M}, or to
take into account heating due to nuclear reactions in the core of the
star
\citep[e.g.][]{DSB06,2021A&A...651A..66K,2022ApJ...926...21B}. Most
often the gas is assumed to be fully ionised and to obey the ideal gas
equation $p = \Rgas \rho T$, where $\Rgas = \cP - \cV$ is the gas
constant and $\cP$ and $\cV$ are the specific heat capacities in
constant pressure and volume, respectively \citep[see, however,
  e.g.][for other approaches]{HRY15a,SBBCMS16}.

The viscous force is given by
\begin{eqnarray}
\FFF^{\rm visc} = 2\nu\rho\bm{\mathsf{S}},
\end{eqnarray}
where $\nu$ is the kinematic viscosity and
\begin{eqnarray}
\mathsf{S}_{ij}\, = \,\onehalf (U_{i;j} + U_{j;i}) - \onethird \delta_{ij} \bm\nabla\bm\cdot {\bm U},
\end{eqnarray}
is the traceless rate of strain tensor where where the semicolons
denote covariant derivatives \citep[cf.][]{MTBM09}. In principle $\nu$
is a function of density and temperature according to
\cite{1962pfig.book.....S}, but in practice the current simulations
adapt various explicit or implicit large-eddy simulation (LES)
formulations that are geared toward minimizing diffusion and
maximising numerical stability on a given grid resolution \citep[for a
  review, see e.g.][]{2015SSRv..194...97M}.

Due to the short mean-free path of photons in stellar interiors,
radiation is typically modeled via the diffusion approximation with
\begin{eqnarray}
\bm{\mathcal{F}}^{\rm rad} = -K \bm\nabla T,
\end{eqnarray}
where $K$ is the radiative conductivity which is related to the
opacity $\kappa$ of the matter via
\begin{eqnarray}
K = \frac{16 \sigmaSB T^3}{3 \kappa \rho},
\end{eqnarray}
where $\sigmaSB$ is the Stefan--Boltzmann constant. The radiative
conductivity is often taken to be a fixed function of radius resulting
either from a stellar evolution model
\citep[e.g.][]{BMT11,2022ApJ...933..199H}, or a simpler fixed analytic
prescription producing a qualitatively similar behavior
\citep[e.g.][]{KMCWB13,2018A&A...616A..72W}. Alternatively, $K$ can
also be taken to be dependent on the ambient thermodynamic state via
the Kramers opacity law
\citep[e.g.][]{2020GApFD.114....8K,2021A&A...645A.141V} with
\begin{eqnarray}
\kappa \propto \rho T^{-7/2},
\end{eqnarray}
which allows a non-linear back-reaction of, for example, rotation and
magnetic fields \citep[e.g.][]{2019GApFD.113..149K}. Radiative cooling
and heating can also be included via the heating/cooling term,
\begin{eqnarray}
  \mathcal{H} = - \bm\nabla\bm\cdot\FFF^{\rm rad},
\end{eqnarray}
as is often done in the simulations with the {\sc Rayleigh} code
\citep[e.g.][]{FH16,2022ApJ...928...51B}. Yet another approach is to
relax the thermodynamics toward a fixed reference state using a
Newtonian cooling term as is done in the {\sc Eulag} simulations
\citep[e.g.][]{GCS10,PC14,2018ApJ...863...35S,2019ApJ...880....6G}.

Typical numerical methods need to include some form of subgrid-scale
(SGS) diffusion in the entropy equation to ensure numerical
stability. In some methods, such as those used in the {\sc ASH}, {\sc
  Rayleigh}, and {\sc Pencil Code}, this is explicitly included in a
term that is proportional to the entropy gradient
\citep[e.g.][]{BMT04,KMCWB13,2020ApJ...892..106M}
\begin{eqnarray}
\bm{\mathcal{F}}^{\rm SGS} = - \chiSGS \rho T \bm\nabla s,
\end{eqnarray}
where $\chiSGS$ is the SGS thermal diffusivity that is responsible for
turbulent diffusion at unresolved scales. This definition implicitly
assumes that $ds/dr < 0$, that is, that the turbulent diffusion is due
to unresolved Schwarzschild unstable convection, and the SGS term
contributes to a positive (outward) energy flux. Often it is
advantageous to decouple the SGS diffusion from the mean
stratification such that the SGS diffusion is applied not on the total
entropy $s$ but, for example, to deviations from the spherically
symmetric mean state $s' = s - \mean{s}$, where the overbar denotes
suitable averaging, typically over the horizontal directions. This
leads to a vanishing mean SGS flux, $\mean{\bm{\mathcal{F}}}^{\rm SGS}
\approx 0$. This is advantageous if part of the convection zone is
weakly stably stratified, or a stably stratified radiative layer is
taken into account below the convection zone
\citep[e.g.][]{BMT11,2020GApFD.114....8K}. An alternative approach is
to include SGS effects implicitly such that the effective diffusion at
small scales is determined by the numerical scheme itself. This is
done in, for example, the {\sc Eulag} \citep[e.g.][]{GCS10} and {\sc
  R2D2} codes \citep[e.g.][]{HRY14}.

In practice, all of the diffusion coefficients in the simulations are
much larger than their counterparts in stars, e.g.\ such that $\nu \gg
\nu_\star$, $\eta \gg \eta_\star$, where the subscript $\star$ refers
to stellar values. Furthermore, the radiative diffusivity $\chi =
K/(\cP\rho)$ is also practically always much smaller than $\chiSGS$
\citep[see Appendix~A of][]{2017A&A...599A...4K}. Therefore all of the
current simulations need to be understood as large-eddy simulations
(LES). Yet, most of them do not consider the non-dissipative
contribution of the unresolved small scales. Furthermore, some models
\citep[e.g.][]{SBBCMS16,2022ApJ...933..199H} dispense with the
physical diffusion terms completely in order to minimize the diffusion
on resolved scales while exerting diffusion only at scales near the
grid scale.

\subsection{Dimensionless parameters and diagnostics}
\label{sec:dimpar}

A number of non-dimensional parameters arise in the analysis of the
MHD equations and which define the simulations. These include the
Rayleigh number which describes the efficiency of convection
\begin{eqnarray}
\Ra = \frac{gd^4}{\nu \chi} \left(- \frac{1}{\cP}\frac{{\rm d}s}{{\rm d}r} \right),
\end{eqnarray}
where $d$ is a length scale; typically taken to be the shell
thickness, the thermal and magnetic Prandtl numbers describe the
relative importance of various diffusion terms:
\begin{eqnarray}
\Pra = \frac{\nu}{\chi},\ \PrM = \frac{\nu}{\eta},
\end{eqnarray}
and the Taylor number
\begin{eqnarray}
\Ta = \frac{4\Omega_0 d^4}{\nu^2},
\end{eqnarray}
which measures the strength of rotation. The latter is related to the
Ekman number via $\Ek = 2 \Ta^{-1/2}$.  Most often the relevant
thermal Prandtl number is based on the SGS diffusion
\begin{eqnarray}
\PraSGS = \frac{\nu}{\chiSGS},
\end{eqnarray}
because $\chiSGS \gg \chi$. For completeness, the viscosity $\nu$ used
in simulations is also an effective or SGS viscosity because it is
always much larger than the real physical value. However, it has the
same functional form as the physical viscosity whereas a term
corresponding to the SGS entropy diffusion does not appear in the
original equations. Additionally the geometry and the resulting
density stratification are input parameters of the models, along with
the boundary conditions applied to the various quantities.

The most common diagnostic parameters used to describe the simulations
include the Reynolds and P\'eclet numbers
\begin{eqnarray}
\Rey = \frac{\urms \ell}{\nu},\ \ReM = \frac{\urms \ell}{\eta}=\PrM\Rey,\ \Pe = \frac{\urms \ell}{\chi}=\Pr\Rey,
\end{eqnarray}
where $\urms$ is the rms-velocity and $\ell$ is a length scale, both
of which are outcomes of the simulations. The magnetic Reynolds number
is of particular interest for dynamo simulations due to the
bifurcations related to the excitation of large-scale and small-scale
dynamos (SSD) \citep[e.g.][]{BS05,SSD_ISSI}. The rotational effect on
the flow is measured by the Rossby (inverse Coriolis) number
\begin{eqnarray}
  \Ro = \frac{\urms}{2\Omega_0 \ell} \propto \Co^{-1}.
\end{eqnarray}
An alternative way to define the Rossby number, which automatically
takes the changing length scale into account is
\begin{eqnarray}
\Ro_\omega = \frac{\orms}{2\Omega_0} \propto \Cow^{-1},
\end{eqnarray}
where $\orms$ is the rms-vorticity with
$\bm\omega=\bm\nabla\times\uuu$. Order of magnitude estimates for some
of these parameters in the deep parts of the solar convection zone and
in a core convection zone of a $20M_\odot$ O star in comparison to
current simulations are listed in \Table{tab:dimpar}.

Our discussion above has assumed that the \emph{dimensional} MHD
equations are being solved, in which case these non-dimensional
parameters are diagnostic outputs of the simulations.  An alternative
approach, also employed by many authors \citep[see,
  e.g.][]{2016JFM...808..690G,2020ApJ...902L...3B} is to
non-dimensionalize the governing equations at the beginning; in this
case the various non-dimensional parameters discussed here appear
directly in the equations, and serve as input parameters that specify
the problem.  To illustrate the procedure, suppose we choose to
measure lengths in units of a characteristic length $\ell_{\rm c}$,
times in units of some time $\tau_{\rm c}$, velocities in units of
$u_{\rm c}$, and temperatures in units of $T_{\rm c}$.  That is, we
assume $x = \ell_{\rm c} x_{\rm nd}$, $t = \tau_{\rm c} t_{\rm nd}$,
and so forth, where the ``$\rm nd$" subscript denotes non-dimensional
variables. Then, to take a simple example, the dimensional Boussinesq
momentum equation in the absence of rotation or magnetism in a plane
layer,
\begin{equation}
  \frac{\pd \uuu}{\pd t} + \uuu \bm\cdot \bm\nabla \uuu = -\bm\nabla \varpi+ \nu \bm\nabla^2 \uuu + \alpha g T \hat{\bm z},
\end{equation}
where $\varpi \sim P/\rho$ is a reduced pressure and other symbols
take their usual meanings, would be rewritten as
\begin{equation}
    \frac{u_{\rm c}}{\tau_{\rm c}} \frac{\pd \uuu_{\rm nd}}{\pd t_{\rm nd}} + \frac{u_{\rm c}^2}{\ell_{\rm c}} \uuu_{\rm nd} \bm\cdot \bm\nabla_{\rm nd} \uuu_{\rm nd} = -\frac{\varpi_{\rm c}}{\ell_{\rm c}} \bm\nabla_{\rm nd} \varpi_{\rm nd} + \frac{u_{\rm c}}{\ell_{\rm c}^2} \nu \bm\nabla_{\rm nd}^2 \uuu_{\rm nd} + \alpha g T_c T_{\rm nd} \hat{\bm z},
\end{equation}
where we have retained $\rm nd$ subscripts on all non-dimensional
quantities (including the spatial and temporal derivatives), and where
$\hat{\bm z}$ is the vertical unit vector. Many choices of the
characteristic scales $\tau_{\rm c}$, $\ell_{\rm c}$, etc., are
possible, and in general these will each yield slightly different
forms of the non-dimensional equations. A common choice is to measure
lengths in units of the convection zone thickness ($\equiv L$), times
in units of a thermal diffusion time across that length ($\tau_{\rm c}
=L^2/\chi$), and to take $u_{\rm c} = L/\tau_{\rm c}$ for consistency;
upon substitution (and simplification) we then find
\begin{equation}
      \frac{\pd \uuu_{\rm nd}}{\pd t_{\rm nd}} + \uuu_{\rm nd} \bm\cdot \bm\nabla_{\rm nd} \uuu_{\rm nd} = -\frac{1}{\Ma^2} \bm\nabla_{\rm nd} \varpi_{\rm nd} + \Pra \bm\nabla_{\rm nd}^2 \uuu_{\rm nd} + \Ra \Pra T_{\rm nd} \hat{\bm z},
\end{equation}
now involving $\Ma^2 = (P_{\rm c}/\rho_{\rm c})/u_{\rm c}^2$, $\Pra =
\nu/\chi$, and $\Ra =g \alpha T_{\rm c} L^3/(\nu \chi)$ (versions of
the Mach, Prandtl, and Rayleigh numbers) as \emph{input} parameters.

An advantage of this approach is that it is easier to avoid
inadvertently ``running the same simulation twice" -- that is,
conducting calculations with different luminosities, rotation rates,
etc., that are nonetheless functionally equivalent (because they have
the same governing non-dimensional parameters). On the other hand, it
is sometimes difficult to ``re-dimensionalize" such calculations and
so make contact with any given astrophysical object; for illustrations
of the procedure and its ambiguities, see discussions in
\citet{2011Icar..216..120J} and \citet{Yadav+16}. We generally adopt
the ``dimensional" view throughout the remainder of this review.

\begin{table}[t!]
\begin{center}
\begin{minipage}{\textwidth}
\caption{Orders of magnitude of some dimensionless parameters in the
  main sequence phase of the Sun in the bulk of convection zone and in
  a core convection zone of a $20M_\odot$ O9 star. Typical values from
  current global 3D simulations of stellar convection and dynamos are
  listed in the last column.}\label{tab:dimpar}
\begin{tabular*}{\textwidth}{@{\extracolsep{\fill}}lccc@{\extracolsep{\fill}}}
\toprule%
Parameter & Sun ($M_\odot$) & O9 ($20M_\odot$) & Simulations \\
\midrule
$\Ra$        & $10^{20}$ & $10^{24}$ & $10^9$  \\
$\Pra$       & $10^{-6}$ & $10^{-5}$ & $10^{-1} \ldots 10$ \\
$\PrM$       & $10^{-3}$ &   $10^3$  & $10^{-1} \ldots 10$ \\
$\Rey$       & $10^{13}$ & $10^{11}$ & $10^{4}$ \\
$\Pe$        &  $10^7$   &  $10^{6}$ & $10^{4}$ \\
$\ReM$       & $10^{10}$ & $10^{14}$ & $10^{4}$ \\
$\Delta\rho$ &  $10^6$   &     $3$   &  $10^2$ \\
$\Ro$        & $0.1\ldots1$ & $1$\footnotemark[1] & $10^{-2} \ldots 10^3$ \\
\botrule
\end{tabular*}
\footnotetext{Note: Solar values are from \cite{O03} and
  \cite{2020RvMP...92d1001S} whereas the values for the O9 star are
  from \cite{2022ApJS..262...19J}. $\Delta\rho$ is the ratio of the
  fluid density between the bottom and top of the convection zone. We
  note that in \cite{2019ApJ...876...83A} the Prandtl numbers for the
  O9 star are somewhat lower, i.e., $\Pra = 10^{-6}$ and $\PrM =
  0.1$.}  \footnotetext[1]{Estimated using $\Ro = \Prot/\tauconv$
  where $\tauconv$ was taken from Fig.~74 of
  \cite{2022ApJS..262...19J} and the solar rotation period
  $P_\odot=27$~days was used as a reference value for $\Prot$.}
\end{minipage}
\end{center}
\end{table}

\subsection{Relevant time and length scales in stars}

The structure of a star is determined by its mass $M$, luminosity $L$,
chemical composition $\mu$, and rotation rate $\Omega_0$, the latter
corresponding to its age. This information, along with material
properties such as viscosity, opacity, equation of state, and nuclear
energy production rate is in principle enough to construct a
time-dependent model of the evolution of the star
\citep[e.g.][]{2012sse..book.....K}. However, in practice the
evolution of main-sequence stars occurs over the nuclear timescale
$\taun$ which is of the order of $\taun \approx 10^{10}$~yr for the
Sun, which is much longer than what can be covered in any 3D dynamo
simulation. Chemical evolution due to nuclear reactions occurs also in
this timescale and therefore the solar and stellar dynamo simulations
assume that the stellar structure is given and fixed in the course of
the simulations \citep[see, however][]{Emeriau+17}. By the same token,
the gravitational potential is assumed to be fixed and spherically
symmetric. Rotational evolution of stars also happens on timescales of
$10^8$ to $10^9$ years
\citep[e.g.][]{1972ApJ...171..565S,2003ApJ...586..464B,Gallet+13} such
that in 3D simulations the rotation rate of the star is assumed to be
fixed.  There is an ongoing debate based observational results
suggesting magnetic braking slows down around the solar age which
might be due to a transition to anti-solar differential rotation and a
corresponding change in the dynamo
\citep[e.g.][]{2016Natur.529..181V}.

In general, the thermal evolution of the star in 3D simulations still
occurs on a Kelvin--Helmholtz timescale
\begin{eqnarray}
\tauKH = \frac{GM^2}{2RL},
\end{eqnarray}
where $G$ is the gravitational constant. The Kelvin-Helmholtz time for
the solar convection zone is of the order of $10^5$ years which is
still about two orders of magnitude longer than the longest global 3D
simulations to date \citep{PC14,KKOBWKP16}. Various ways to overcome
or circumvent this issue are discussed below in
\Sec{sec:simstrat}. The final timescale related to stellar structure
is the free-fall or acoustic timescale $\tau_{\rm ac} =
\sqrt{R^3/GM}$, which is of the order of 30 minutes for the Sun.

In terms of global dynamos the most important timescales are the
rotation period $\Prot$ and the convective turnover time,
\begin{eqnarray}
\tauconv = \frac{\ell}{\uconv},
\end{eqnarray}
where $\ell$ is the convective length scale and $\uconv$ a suitably
averaged convective velocity. The convective turnover time $\tauconv$
can be estimated from solar surface observation where granules
overturn on a timescale of a few minutes. Knowledge of $\tauconv$ in
deeper layers relies heavily on theoretical estimates, for example,
from mixing length models \citep[e.g.][]{BV58}. These assume that the
length scale is proportional to the pressure scale height. At the same
time, convective velocities decrease such that $\tauconv$ is of the
order of a month near the base of the solar convection zone
\citep[e.g.][]{Stix02}. Stellar observations indicate that dynamo
efficiency of stars is related to the Rossby number (see,
\Sec{sec:sunintime})
\begin{eqnarray}
\Ro = \frac{\Prot}{\tauconv}.
\end{eqnarray}
The Rossby number is the only non-dimensional diagnostic that the
simulations can capture relatively accurately; see \Table{tab:dimpar}
and \Sec{sec:simstrat}.

Another timescale that the simulations need to capture is the activity
cycle period $\taucyc$ which is 22 years for the Sun, and which varies
between years to decades for stars other than the Sun
\citep[e.g.][]{1995ApJ...438..269B,2009AJ....138..312H,2018A&A...619A...6O}. It
is, however, practically always necessary to run simulations
considerably longer because establishing the global dynamo and
reaching the final saturated dynamo mode takes typically significantly
longer \citep[e.g.][]{KMCWB13,2020ApJ...892..106M}. Taking the thermal
relaxation also into account, the integration times are typically at
least an order of magnitude longer than the cycles established. The
necessity to run such long times is one of the major limiting factors
in the quest to reach astrophysically relevant parameter regimes.

In principle the relevant length scales vary between the depth of the
convection zone $\Delta R$ (or the radius of the star $R$ for fully
convective stars), and the Kolmogorov length scale $\ell_\nu$ where
kinetic energy is dissipated into heat due to viscosity (or to the
magnetic dissipation scale $\ell_\eta$ in cases where $\PrM > 1$).
According to the Kolmogorov turbulence phenomenology
\citep[e.g.][]{1995turb.book.....F}, $\ell_\nu$ can be estimated from
the Reynolds numbers at the integral ($L$) and Kolmogorov scales,
\begin{eqnarray}
\ell_\nu = L \left( \frac{\Rey_L}{\Rey_{\ell_\nu}} \right)^{-3/4}.
\label{equ:ellnu}
\end{eqnarray}
For the Sun, $L=\Delta R = 2\cdot10^8$~m and $\Rey_L \gtrsim 10^{12}$
\citep[e.g.][]{O03,2022ApJS..262...19J} and $\Rey_{\ell_\nu} = 1$ by
definition. These estimates yield an upper limit of order of magnitude
$\ell_\nu \approx 0.1$~m near the base of the solar convection
zone. More detailed calculations yield values between $0.01$~m
\citep{2017LRCA....3....1K} and $0.06$~m
\citep{2020RvMP...92d1001S}. Furthermore, the dissipation scales of
magnetic fields and temperature fluctuations can be estimated from
$\ell_\eta = \ell_\nu \PrM^{-3/4}$, and $\ell_\chi = \ell_\nu
\Pra^{-3/4}$, respectively. Even though $\PrM, \Pra \ll 1$, these
scales are also very small in comparison to the depth of the
convection zone or the radius of the star. As will be discussed below,
all of the scales where physical diffusion occurs are several orders
of magnitude smaller than what can be achieved in any current or
foreseeable simulations \citep[see also][]{2017LRCA....3....1K}.
 
\begin{figure*}[t]
\begin{centering}
  \includegraphics[width=0.99\textwidth]{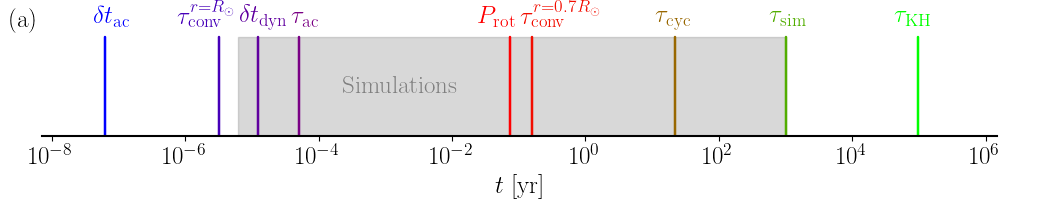}
  \includegraphics[width=0.99\textwidth]{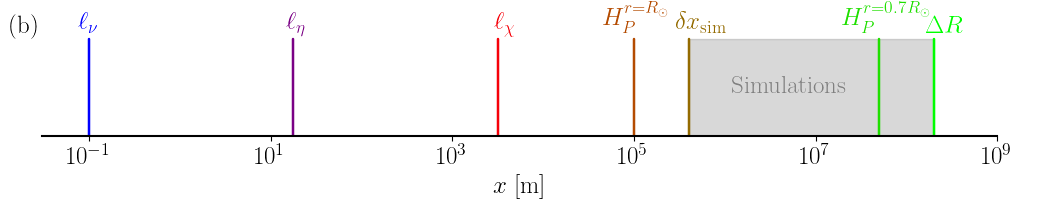}
\caption{Orders of magnitudes of time (a) and length (b) scales in the
  deep solar convection zone. The gray shaded area indicates regions
  accessible to a typical current global simulation.}
\label{fig:scales}
\end{centering}
\end{figure*}

Another length scale that plays an important role is the pressure
scale height $\HP = - dr/d\ln p$ which is related to the vertical
scale of convection cells. Near the surface $\HP$ is of the order of
$100$~km in the photosphere of the Sun. On the other hand, at the base
of the convection zone $\HP \approx 5\cdot10^4$~km. This reflects the
fact that near the surface the pressure and density decrease very
rapidly and that capturing both the deep and photospheric convection
in a single model is therefore very challenging. In total the solar
convection zone encompasses more than 20 pressure scale heights. This
translates to a density difference of $10^6$ between the photosphere
and the base of the convection zone. Estimates of the relevant
temporal and length scales in the deep parts of the solar convection
zone are summarized in \Fig{fig:scales}.

\section{Numerical approach to stellar dynamos}
\label{sec:numapp}

\subsection{Simulation strategy}
\label{sec:simstrat}

The main difficulty in solar and stellar dynamo simulations is that it
is not feasible to match the dimensionless parameters with those of
real stars as seen from the comparison of stellar and simulation
parameters in \Table{tab:dimpar}. The only general exception to this
is the Rossby number but even there we face the situation that only
some of the convective scales in stars are strongly affected by
rotation. For example, in the Sun the near-surface layers are
practically unaffected by rotation whereas at the base of the
convection zone the Rossby number is of the order of 0.1
\citep[e.g.][]{O03}. This is to be contrasted with the Earth's dynamo
where the magnetic Reynolds number is $\ReM\approx10^3$, which is
within reach of current simulations
\citep[e.g.][]{2021GeoJI.225.1854A} and where all convective scales
are strongly rotationally constrained ($\Ro\approx10^{-6}$); see,
e.g., \cite{2013RPPh...76i6801R}.

A path that stellar dynamo simulations often follow to approach
physically relevant regimes is to assume a fixed convective Rossby
number \citep{Gi77}, given by
\begin{eqnarray}
  \Roc = \left(\frac{\Ra}{\Pra \Ta}\right)^{1/2}.
\end{eqnarray}
Here, the stellar luminosity fixes the level of driving through the
Rayleigh number, and the stellar rotation rate is fixed by
observations. Using typical estimates for $\Ra$, $\Pra$, and $\Ta$ for
the Sun \citep[][see also \Table{tab:dimpar}]{O03} we arrive at $\Roc
\approx 0.1 \ldots 1$. In simulations, the (SGS) Prandtl number
$\PraSGS=\nu/\chiSGS$ is often fixed and changing the diffusivities
$\nu$ and $\chiSGS$ leads to $\Ra \propto \nu^{-2}$, $\Ta \propto
\nu^{-2}$ and $\Roc = \rm{const.}$ An obvious limitation is that the
Prandtl number in simulations is typically close to unity whereas in
stars $\Pra \ll 1$
\citep[e.g.][]{2019ApJ...876...83A,2020RvMP...92d1001S,2022ApJS..262...19J}. A
similar argument applies to $\PrM$ in late-type stars whereas in the
core convection zones of massive O and B stars $\PrM\gg1$
\citep[e.g.][]{2016ApJ...829...92A}. Furthermore, in iLES models the
values of the dimensionless parameters are typically unknown and not
precisely controllable, although it is often possible to determine
these \emph{a posteriori}
\citep[e.g.][]{SBBCMS16,2022ApJ...933..199H}. Nevertheless, the
strategy in both LES and iLES models is to try to capture the stellar
Rossby number with unrealistic Prandtl numbers and to resolve enough
scales in an effort to reach sufficiently high $\Rey$, $\ReM$, and
$\Pe$ such that the large scale results are no longer
affected. However, it is still questionable whether such a regime has
been reached even in the highest resolution simulations to date
\citep[e.g.][]{2022ApJ...933..199H,Guerrero+22}.

\subsection{Limitations of current numerical simulations}
\label{sec:limitations}

The challenge of doing direct numerical simulations (DNS) of stars is
illustrated by considering the solar convection zone where the fluid
Reynolds number is of the order of at least $10^{12}$
\citep[e.g.][]{O03,2022ApJS..262...19J}. With this estimate and
\Eq{equ:ellnu}, the ratio of the system scale $L$, here taken to be
the depth of the solar convection zone or $200$~Mm, to the Kolmogorov
scale $\ell_\nu$, is
\begin{eqnarray}
\frac{L}{\ell_\nu} = \left( \frac{\Rey_L}{\Rey_{\ell_\nu}} \right)^{3/4}.
\end{eqnarray}
With $\Rey_{\ell_\nu}=1$, we obtain $L/\ell_\nu =10^9$. This ratio
gives the order of magnitude of grid points that is required to
capture all of the physically relevant scales in the solar convection
zone. Thus a direct 3D simulation requires $10^{27}$ grid points. A
somewhat lower, but still unattainable, number was reported in
\cite{CS86}.

Current state-of-the-art global simulations have of the order of
$10^{10}$ grid points and are run on a few times $10^4$ CPU
cores. Assuming ideal weak scaling, where the computation time remains
constant when the number of CPUs is increased proportional to the grid
size, a DNS of the solar convection zone requires $10^{21}$ CPU
cores. Using a current 96-core AMD Epyc\texttrademark\ 9654 CPU with
360~W thermal design power as a
reference\footnote{\href{https://www.amd.com/en/products/cpu/amd-epyc-9654}{https://www.amd.com/en/products/cpu/amd-epyc-9654}},
gives a total power consumption of $3.8\cdot10^{21}$~W, corresponding
roughly to a M9V main-sequence red dwarf. It is clear that such power
is neither available nor meaningful to be spent. Although reaching an
asymptotic regime where the large-scale dynamics are unaffected by the
addition of further small scales is very likely possible at a
significantly lower resolution, it is clear that even the highest
resolution current simulations are not there yet
\citep[e.g.][]{2017A&A...599A...4K,2022ApJ...933..199H}.

Furthermore, the timestep in such hypothetical DNS of the solar
convection zone is of the order of $\delta t \approx \ell_\nu/{\rm
  max}(c_{\rm sig}^{\rm max})$, where $c_{\rm sig}^{\rm max}$ is the
maximum signal propagation speed. In anelastic models this is set by
the maximum flow velocity which is of the order of 1~km s$^{-1}$,
whereas in the fully compressible case this is the sound speed $\cs$,
which at the base of the convection zone is around
200~km~s$^{-1}$. This gives $\delta t = \delta t_{\rm dyn} \approx 2
\times 10^{-4}$~s for anelastic and $\delta t = \delta t_{\rm ac}
\approx 10^{-6}$~s for fully compressible models. In practice, the
resolution is much lower and corresponding estimates for a
high-resolution global simulation with 500 uniformly spaced grid
points in radius gives $\delta t_{\rm ac} \approx 2$~s and $\delta
t_{\rm dyn} \approx 10$~minutes. The latter is still longer than the
convective turnover time near the surface of the Sun, where $\tau_{\rm
  conv}^{r=R_\odot} \approx 1$~minute. The surface of the Sun is
extremely challenging to be taken into account in a global model due
to a combination of very small length scales and short time scales and
the Mach number approaching unity. Therefore a full Sun simulation
requires a numerical scheme capable of dealing with practically all
Mach numbers and multiscale convection. Furthermore, the boundary
region where radiative cooling takes place near the surface is
extremely thin, around 10~km, in comparison to the depth of the
convection zone \citep{2017LRCA....3....1K}. Typically simulations
either do not reach all the way to the photosphere, or consider a
shell reaching to $R=R_\odot$ but with a much lower density
stratification than in the Sun, and the boundary layer near the
surface is made artificially thicker to resolve it numerically.

Another constraint arises due to a widening discrepancy of the
timescales involved when resolution is increased: as was discussed
earlier, a simulation of the Sun needs to cover at least a solar cycle
or preferably several cycles to be considered viable such that the
simulated time $\tau_{\rm sim} \gtrsim \taucyc$. For the sake of
argument, an acceptable maximum wall-clock time that a simulation is
permitted to run to is taken to be a year. This requires that the star
in the simulation has to evolve at least 22 times faster than in real
time. However, when the grid resolution is increased, the timestep in
explicit time-stepping methods decreases in proportion with the grid
spacing $\delta x$, and the computational cost of simulation increases
with $\delta x^4$. Even if the numerical scheme has ideal weak
scaling, the time to solution doubles every time the resolution is
doubled, which typically cannot be avoided. This poses stringent
constraints on either the length, or the grid resolution, of
simulations targeting global stellar dynamos. The timestep constraints
can, to a certain degree, be alleviated by the use of implicit time
stepping methods \citep[e.g.][]{2011A&A...531A..86V} or by the use of
local subgrids and timesteps \citep[e.g.][]{2022arXiv221109564P}.

A futher complication arises due to the thermal relaxation. In
anelastic models, where the real stellar luminosity is often used, the
Kelvin-Helmholtz time is much longer than the integration times of
simulations. However, this is a worst-case scenario because deep
stellar convection zones are nearly adiabatic which is exploited in
the simulation setups. Thermal relaxation can still take a
prohibitively long time if a stably stratified radiative layer is
retained below the convection zone. This issue is sometimes alleviated
by adjusting the radiative conductivity in the overshoot layer below
the convection zone; see e.g.\ \cite{2017ApJ...836..192B}. However,
this can lead to over- or underestimation of convective overshooting
depending when and how such adjustments are made
\citep{2019A&A...631A.122K}. Another possibility is to adjust the
thermodynamic state and fluctuations recursively toward an equilibrium
solution \cite{2018PhRvF...3h3502A,2020PhRvF...5h3501A}, although this
method has yet to gain widespread adoption in compressible or global
3D simulations.

In fully compressible simulations the timestep would be very short
because it is determined by the sound speed at the base of the
convection zone. This has been circumvented by the reduced sound speed
technique (RSST) where the sound speed is artificially lowered such
that the timestep issue is alleviated \citep[e.g.][]{HRY14}. Another
approach is the enhanced luminosity method (ELM) where a luminosity
that is much higher than in real stars is used
\citep[e.g.][]{KMCWB13,2020GApFD.114....8K}, leading to a higher Mach
number and therefore a diminished gap between the acoustic and
dynamical timescales, as well as a correspondingly shorter
Kelvin-Helmholtz time. Given that the luminosity enhancement is
sufficiently large, it is possible to resolve the Kelvin-Helmholtz
timescale using fully compressible MHD equations
\citep[e.g.][]{2023A&A...669A..98K}. The cost of this method is that
in addition to higher flow velocities, also the thermodynamic
fluctuations are enhanced, and a direct comparison with observations
requires the use of scaling relations. Furthermore, to achieve the
same Rossby number as in a real star with realistic luminosity, the
rotation rate has to be increased in proportion to the Mach number,
which would lead to unrealistically large centrifugal force
\citep[e.g.][]{2022A&A...667A.164N}.

\subsection{Numerical methods and codes}\label{subsec:codes}

There are a variety of codes solving the MHD equations in spherical
shells and targeting solar and stellar global dynamos. The first and
still popular approach is to adopt the anelastic approximation where
the sound waves are filtered out by neglecting the time derivative in
continuity equation. Then it is convenient to use spherical harmonics
to solve for the horizontal dynamics whereas the vertical
discretisation is often done with finite differences or Chebychev
polynomials. Codes using this approach include {\sc ASH}
\citep[e.g.][]{Clune,BMT04,2011Icar..216..120J,2022ApJ...926...21B},
{\sc Rayleigh} \citep[][]{featherstone_et_al_2022}, and {\sc MagIC}
\citep[e.g.][]{2012Icar..219..428G}.  {\sc Eulag} is another anelastic
code but instead of spherical harmonics, it relies on a second-order
accurate multidimensional positive-definite advection transport
algorithm (MPDATA) and implicit time stepping \citep[e.g.][]{SC13}.

Another popular technique is to use the fully compressible formulation
and using some flavour of finite difference methods. This typically
leads to coordinate singularities at the axis and at the centre of the
star, which are circumvented by either omitting regions near the axis
\citep[spherical wedge, cf.][]{KMB12,MMK15}, using partially
overlapping grids \citep[yin-yang grid cf.][]{HRY15a}, or embedding a
spherical star into a Cartesian cube \citep[star-in-a-box model,
  cf.][]{DSB06,2021A&A...651A..66K}. The Mach number issue of fully
compressible simulations is dealt with by RSST and ELM methods that
were discussed above. Codes using fully compressible formulation
include the {\sc Pencil Code} \citep{2021JOSS....6.2807P}, {\sc R2D2}
\citep[][]{HRY15a}, and the code used in \cite{MMK15}. Further methods
include the {\sc Dedalus} framework which uses spectral methods and is
capable of solving incompressible, anelastic, and fully compressible
equations in varying geometries
\citep{2020ApJ...902L...3B,2020PhRvR...2b3068B,2022arXiv220400002A},
and the {\sc Dispatch} framework where various solver for compressible
flows are possible and which uses local subdomains and timesteps
\citep{2018MNRAS.477..624N,2022arXiv221109564P}.

\section{Relevant solar and stellar observations}\label{sec:observations}

The dynamo simulations discussed in this review aim, ultimately, to
capture the flows and magnetism occurring in real stars. Here, we
briefly describe some of the most pertinent observational constraints
on these processes, which also serve to motivate and guide our work.
Perhaps the most obvious characteristics that any dynamo simulation
would hope to match are the Sun's observed differential rotation and
its periodic cycle of magnetic activity.

\begin{figure*}[t]
\begin{centering}
  \includegraphics[width=0.5\textwidth]{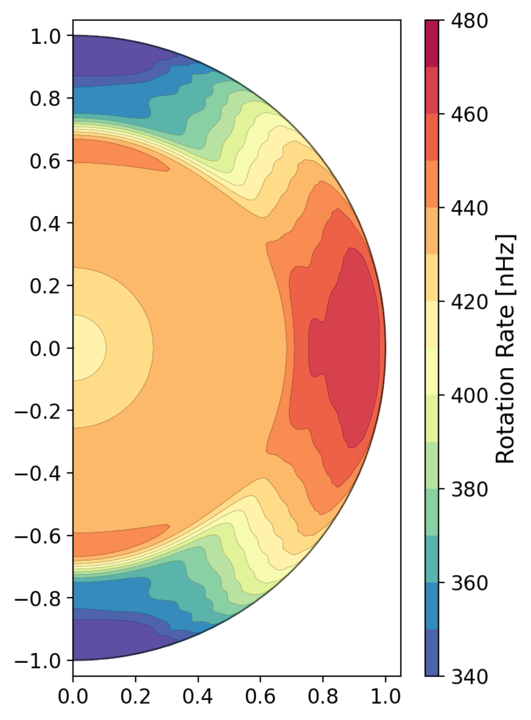}
\caption{Solar differential rotation profile $\Omega/2\pi$ as a
  function of radius and latitude as inverted from helioseismology
  using the first 6 years of HMI 72-day analysis. Reddish colors
  indicating fast rotation and blueish tones slow rotation (adapted
  from \cite{2018SoPh..293...29L} using the data archived at
  \href{http://jsoc.stanford.edu/HMI/Global\_products.html}{http://jsoc.stanford.edu/HMI/Global\_products.html}).}
\label{fig:rot}
\end{centering}
\end{figure*}

In Figure \ref{fig:rot}, we sample the solar interior rotation profile
as revealed by helioseismology \citep[e.g.][]{Schouea98}. The Sun's
surface differential rotation -- with a fast equator and a slow pole
-- imprints through the convection zone, with nearly solid-body
rotation in the portions of the radiative zone below that are
accessible to these global-scale inversions. There are two prominent
shear layers -- the tachocline near the base of the convection zone
and the near-surface shear layer (NSSL) in the upper portions of the
convection zone. The apparent width of the tachocline in this
representation reflects the width of the inversion kernels used; its
true width is thought to be narrower.  Note, too, that although the
shear within the convection zone has not spread through the radiative
interior, the \emph{average} rotation rate of the interior is
commensurate with the average rate of the envelope; since the Sun is
continuously losing angular momentum via its magnetized wind, and so
spun more rapidly in the past, this observation implies some level of
coupling between the two regions
\citep{1989ApJ...338..528G,1991ApJ...376..204M,1992A&A...265..106S,1998Natur.394..755G,BMT11,2015ApJ...799L..23M}.

Ideally, a simulation would self-consistently capture at least a few
key attributes of this profile: e.g., the overall pole-to-equator
shear within the convection zone; the fact that isocontours of
$\Omega$ are more nearly aligned with radius than they are with the
rotation axis, in evident tension with the Taylor-Proudman theorem
\citep[e.g.][]{MBT06}; and the existence and properties of both the
NSSL and the tachocline. In practice each of these remain a challenge,
though as discussed later in this review (\Sec{sec:cursun}), the
latest 3D MHD global simulations of solar interior dynamics and
angular momentum transport do manage to capture many of these elements
without undue tinkering.

\begin{figure*}[t]
\begin{centering}
  \includegraphics[width=0.95\textwidth]{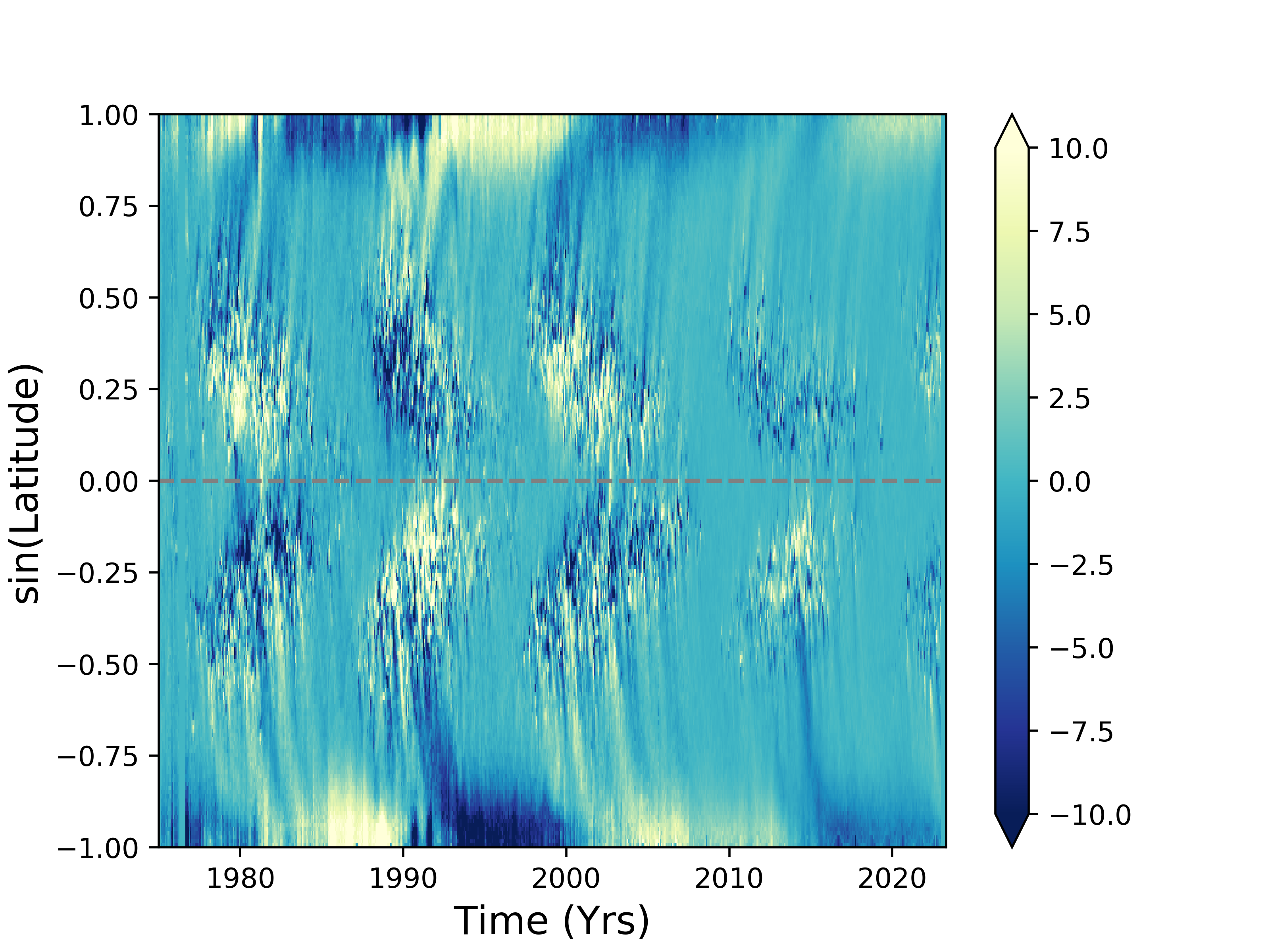}
\caption{Solar butterfly diagram of the line of sight surface magnetic
field up to Carrington rotation 2265 \citep{2020IAUS..354..138B}.}
\label{fig:btfy}
\end{centering}
\end{figure*}

The most striking observational constraints on the magnetism involve
its systematic evolution with space and time, as sampled in the famous
``butterfly diagram.''  An example is provided in Figure
\ref{fig:btfy}, which shows the longitudinally-averaged line of sight
component of the magnetic field for every Carrington rotation since
1975 \citep[based on Wilcox, GONG and Solis synoptic map
  data;][]{2020IAUS..354..138B}. Strong fields emerge at mid-latitudes
and then, over the course of roughly 11 years, appear progressively
nearer the equator; the polarity of these emergent fields is the same
for most of the low-latitude events in the Northern hemisphere, and
opposite to that in the Southern; the overall polarity of the field
flips at the end of each 11-year period. There is also a prominent
polar branch of activity, which is at its strongest when the
equatorial branch is at its ebb; the polarity of this polar branch
matches that of the \emph{following} equatorward branch. The polarity
of the poloidal field thus reverses when active region emergence is at
its maximum. The overall number of sunspots visible over the solar
surface rises and falls over the course of the cycle, in line with the
surface distribution of the strongest fields.

\begin{figure*}[t]
\begin{centering}
\includegraphics[width=0.95\textwidth]{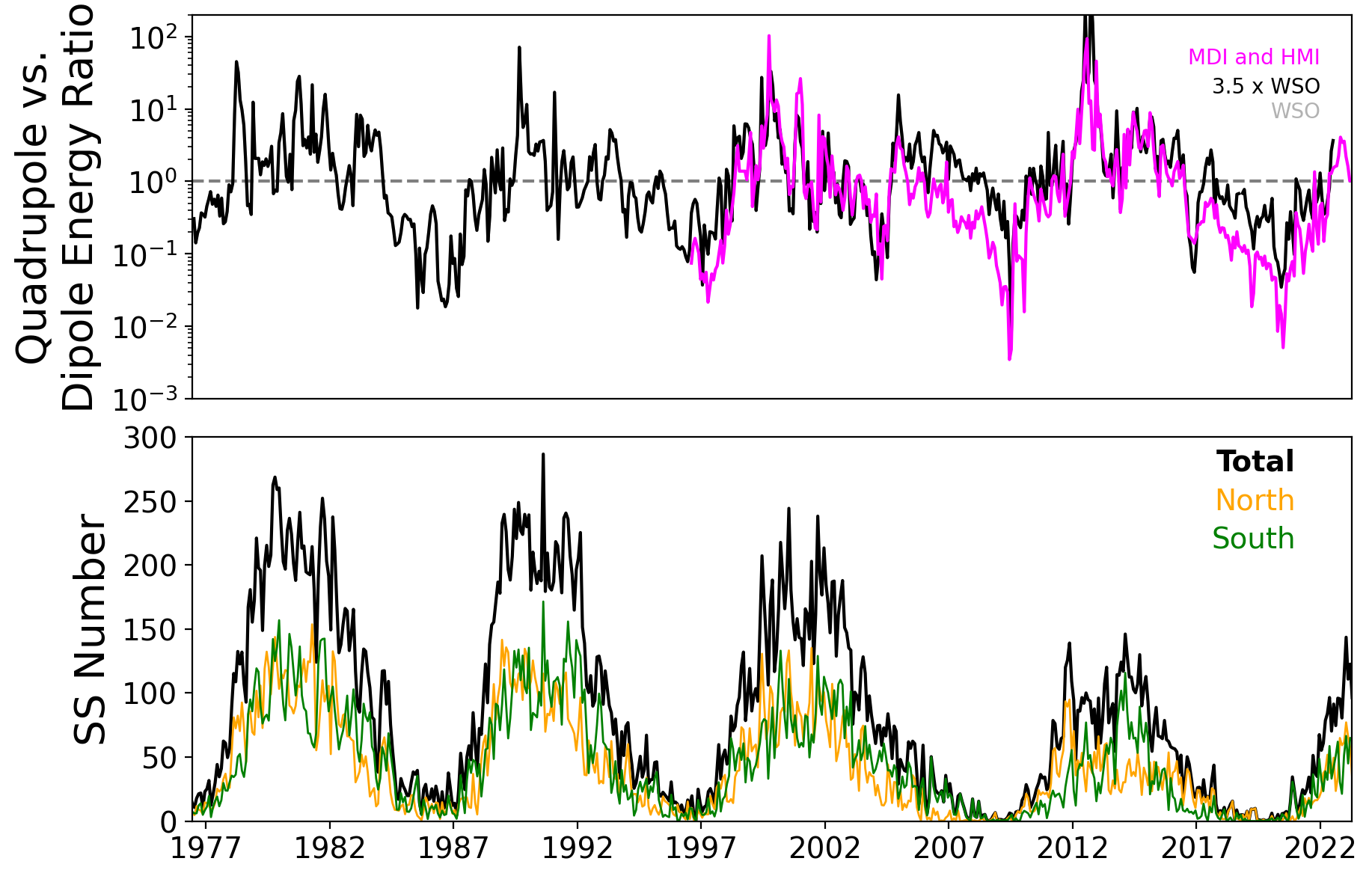}
\caption{Top: Ratio of the solar magnetic dipole and quadrupole energies 
  over the last few cycles (MDI and HMI Data). We note that the
  quadrupole modes dominate during the maximum cycle phase and this
  has already started for cycle 25 in 2022. This strong quadrupolar
  component also explains the time lag between the northern and
  southern hemispheres that can reach up to 18 months. Bottom: Sunspot
  numbers from all data and separately from northern and southern
  hemispheres. Adapted from \cite{2020IAUS..354..138B}.}
\label{fig:DipQuad}
\end{centering}
\end{figure*}

On the whole, the Sun's ordered field exhibits dipole parity
throughout most of the cycle (it is antisymmetric about the equator),
but there are periods near cycle maximum during which the parity is
mostly quadrupolar. The relative amplitudes of the dipolar and
quadrupolar modes are shown in \Fig{fig:DipQuad} over the past few
cycles \cite{2020IAUS..354..138B}. These relations constitute another
powerful constraint on dynamo models. For example, there is evidence
that around the pronounced period of low surface activity known as the
Maunder minimum, the Sun's observed surface activity was predominantly
confined to one hemisphere, indicating different parity relations; the
implications of this finding for dynamos generally (and grand minima
in particular) have been considered by, e.g.,
\cite{1994A&A...288..293S} and many subsequent papers.

Very recently, new constraints have begun to emerge from the study of
inertial and Rossby wave modes that propagate within the convection
zone. These toroidal modes have recently been observed in helioseismic
maps of near-surface horizontal flows obtained by HMI aboard SDO
\citep{2021A&A...652L...6G}; see also \cite{2022NatAs...6..708H} for
another recent detection of solar inertial modes. Though modeling of
these modes is still in its infancy
\citep{2022A&A...666A.135B,2022ApJ...934L...4T} they appear to hold
great promise for constraining aspects of the convection that would be
difficult or impossible to estimate by other means. As a first
application, \cite{2021A&A...652L...6G} illustrate that these modes
constrain the superadiabaticity and turbulent diffusivity of the deep
solar convection zone.

\begin{figure*}[t]
\begin{centering}
  \includegraphics[trim=0 0 3cm 0,clip, width=0.95\textwidth]{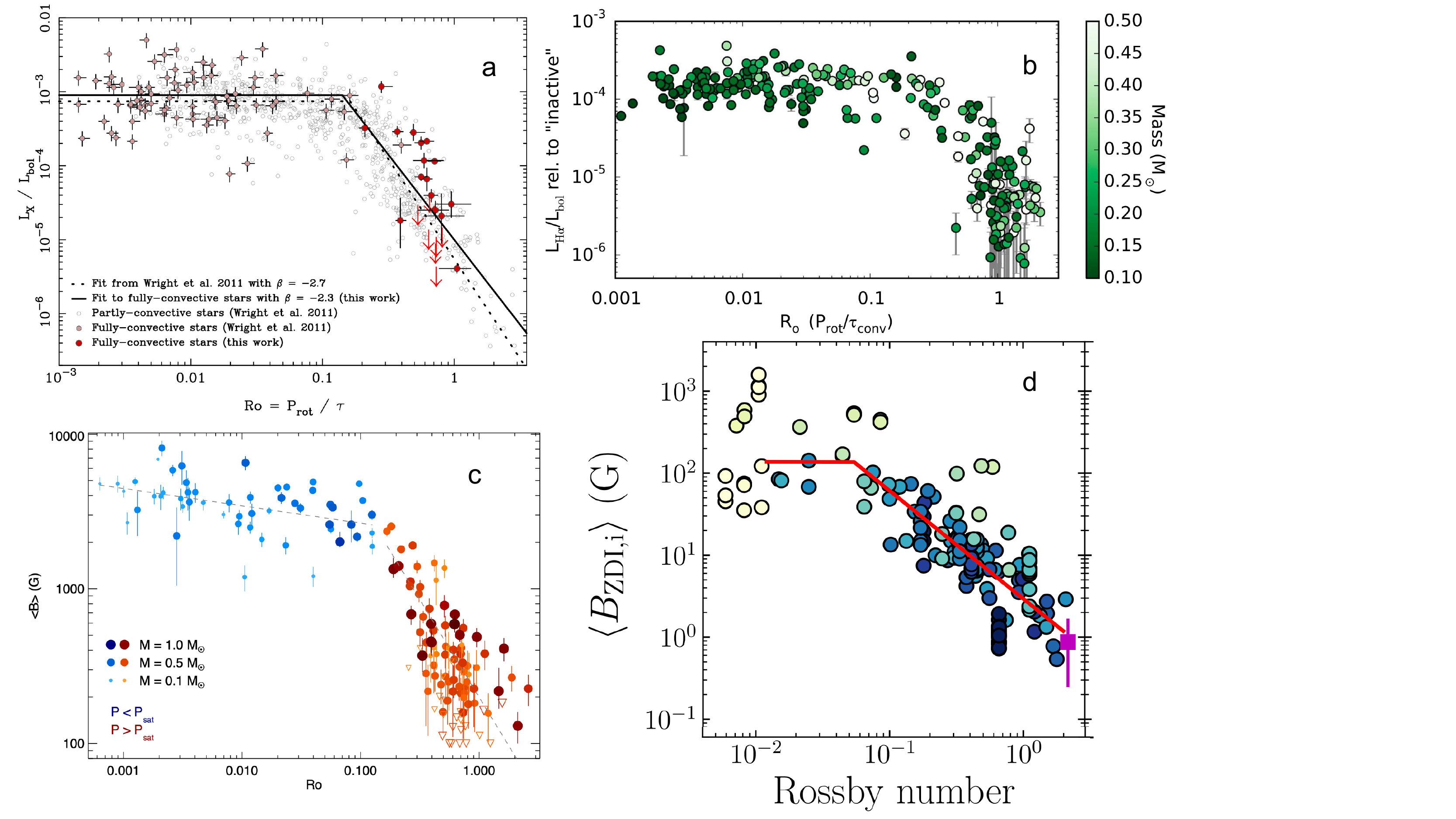}
\caption{Relations between rotation rate (quantified by the Rossby
  number) and magnetic activity, sampled by several different
  observational methods. ($a$) X-ray emission (normalized to the
  bolometric luminosity) for a sample of both fully and partially
  convective stars \citep{2018MNRAS.479.2351W}; ($b$) a measure of
  chromospheric H$\alpha$ emission, normalised to the bolometric
  luminosity and relative to that in inactive stars, in a sample of
  M-dwarfs \citep{2017ApJ...834...85N} (\textcopyright AAS. Reproduced
  with permission); ($c$) average surface magnetic field as measured
  by Zeeman broadening of spectral lines \citep{2022A&A...662A..41R}
  (\textcopyright ESO. Reproduced with permission); ($d$) Zeeman
  Doppler imaging estimates of dipole component of surface magnetic
  field \citep{2019ApJ...886..120S} (\textcopyright AAS. Reproduced
  with permission).}
\label{fig:rotation_activity_composite}
\end{centering}
\end{figure*}

Finally, we turn briefly to observations of other stars. Many aspects
of observational stellar magnetism are treated elsewhere in this
volume, and in other reviews
\citep[e.g.][]{2012LRSP....9....1R,2017LRSP...14....4B} so we note
only a few key constraints that must eventually be matched by
simulations. The most celebrated of such constraint is the
``rotation-activity correlation,'' sampled in
\Fig{fig:rotation_activity_composite}. In stars with convective
envelopes, surface measurements of magnetic activity first increase
with rotation rate, then plateau (``saturate'') at a certain
point. Here we show examples in which activity is measured by coronal
emission \citep{2018MNRAS.479.2351W}, here including both fully
convective and partially-convective stars (panel $a$); by
chromospheric H-$\alpha$ emission (as a fraction of the bolometric
luminosity) in a large sample of M-dwarfs \citep{2017ApJ...834...85N}
(panel $b$); via (panel $c$) measurements of the surface magnetic
field strength as revealed by the Zeeman broadening of spectral lines
\citep[][]{2022A&A...662A..41R}; and (panel $d$) via Zeeman Doppler
Imaging \citep{2019ApJ...886..120S}, here providing an estimate of the
large-scale dipole field observed at the surface.  Typically in these
studies the influence of rotation is characterized via a simple
estimate of the Rossby number where the convective turnover time is
typically based on simple empirical relations that work well for
main-sequence stars \citep[e.g.][]{1984ApJ...287..769N}.  When viewed
in this way, many different types of stars -- including those with and
without a stable radiative region -- appear to exhibit the same basic
relationships. Comparisons of young and evolved main-sequence stars
also suggest a similar level of activity as function of $\Ro$ in terms
of Ca II H\&K emission, provided that the convective turnover time is
an outcome of 1D stellar models \citep{2020NatAs...4..658L}.

The rotational velocity of a solar-like star changes systematically
over time, as it loses angular momentum through a magnetised wind, so
over the course of its life it will trace a variety of positions on
this rotation-activity correlation. Indeed, measurements of rotation
rate are used in ``gyrochronology'' as proxies for age, because many
stars are found to exhibit a common, tight relation between spin rate
and time -- the so-called Skumanich law; see discussions in
\cite{1972ApJ...171..565S}, \cite{1983ApJS...53....1S}, and
\cite{2003ApJ...586..464B}. Lately there have been some indications
that this relationship may break down at late ages
\citep[e.g.][]{2016Natur.529..181V}, which may provide additional
constraints on the dynamo for old stars
\citep[e.g.][]{2017SoPh..292..126M}. In a related vein, there is some
evidence for enhanced stellar activity in a subset of slowly-rotating
stars, which some authors have suggested may be linked to the presence
of strong anti-solar differential rotation
\citep{2018ApJ...855L..22B}.  Global dynamo simulations seem to
capture this effect; see \cite{KKKBOP15}, \cite{2020A&A...642A..66W},
and \cite{2022ApJ...926...21B}.

Together, these observations of the Sun and other stars constitute
powerful constraints that models would hope to satisfy. In the
following sections, we will examine the extent to which they actually
do so.

\section{Simulations of solar and stellar dynamos}\label{sec:results}

\subsection{Convection and dynamo in the current Sun}
\label{sec:cursun}

In addition to the fundamental numerical restrictions discussed above,
simulations of the current Sun are challenging due to the possible
proximity of the transition from solar-like to anti-solar differential
rotation. This transition occurs around $\Ro\approx 1$
\citep[e.g.][]{KMB11,GYMRW14,2017ApJ...836..192B} and current
simulations of the Sun appear to lie close to this in parameter
space. Simulations with the nominal solar luminosity and rotation rate
land predominantly in the anti-solar regime \cite[e.g.][]{KKB14},
which is one of the manifestations of the convective conundrum
\citep{2016AdSpR..58.1475O}, or the lower than expected velocity
amplitudes and Rossby number in the Sun in comparison to theoretical
estimates and simulations, which is discussed in more detail elsewhere
\citep[e.g.][]{2016AnRFM..48..191H}. Therefore simulations targeting
the Sun often resort to lowering the Rossby number artificially to
obtain a solar-like differential rotation profile. This is most often
done by suppressing the convective velocity by enhancing radiative
diffusion \citep[e.g.][]{KKB14,FF14,HRY16,Noraz_thesis}, lowering the
luminosity \citep[e.g.][]{HRY15a,Guerrero+22}, or by increasing the
rotation rate.

The importance of reproducing the solar interior rotation lies in the
fact that the dynamo process crucially relies on flows of various
scales to maintain the observed magnetic field. At the very least, the
large-scale flows are employed by all of the currently predominant
solar dynamo models \citep{2020LRSP...17....4C}. Furthermore, it is
also likely that turbulent effects, such as an $\alpha$ effect due to
helical convection-driven turbulence, are important in the dynamo
process. This highlights the importance of accurate modelling of
convection, essentially necessitating that the solar velocity field
has to be sufficiently well reproduced first before one should expect
success in reproducing the dynamo.  An important step in this is the
identification of the relevant force balances that need to be
reproduced in simulations. Following such an approach, a path in
parameter space may be found that leads to solar-like results with
feasible numerical cost. Such ``path approach" is quite commonly used
now in geodynamo modelling
\citep{2017JFM...813..558A,2021GeoJI.225.1854A}.

The convective conundrum is arguably the greatest obstacle in
achieving the goal of simulating the solar dynamo
successfully. Several ideas have recently been invoked to alleviate
the discrepancy between models and reality. One of these ideas is that
this is a manifestation of rotationally constrained convection in the
interior of the Sun. In this scenario the maximum scale of convection
in the deep parts of the solar convection zone is not giant cells, as
is expected from mixing length models, but it matches instead that of
the supergranulation, which is also detected from surface observations
\citep{FH16b,2021PNAS..11822518V}.  While simulations of rotationally
constrained convection do produce smaller convective scales that
become smaller as rotation becomes more rapid
\citep[e.g.][]{2018A&A...616A.160V}, the main contribution to
differential rotation in such models is still due to giant cell
convection \citep[e.g.][]{2023A&A...669A..98K} or thermal Rossby waves
that have not yet been detected in the Sun.

Another idea that has gained popularity recently is that the deep
parts of the solar convection zone can be weakly stably stratified.
This is thought to result from strong driving of convection in the
near-surface layers, whence plumes of cool low entropy material plough
through the whole convection zone and deep into the stably stratified
layers below. Such idea of \emph{cool entropy rain} was put forward by
\cite{Sp97} and later incorporated into a modified mixing length model
by \cite{Br16}. In the extreme versions of these models only a very
thin layer (down to a few Mm) near the surface of the convection zone
is Schwarzschild unstable and the rest of the convection zone is
weakly subadiabatic and mixed by the entropy rain. Such effects were
explored in 3D simulations by \cite{2018ApJ...859..117N} by means of a
boundary condition consisting of localised cooling patches. Although
non-rotating simulations often find relatively deep subadiabatic
layers
\citep[e.g.][]{2015ApJ...799..142T,2017ApJ...843...52H,2017ApJ...845L..23K},
their effect in global simulations appears to be weak
\citep{2019GApFD.113..149K,2021A&A...645A.141V}.  This could also be
due to the modest resolutions and supercriticality of convection in
those studies.

Furthermore, the influence of the thermal Prandtl number has also
recently been studied. In particular, several studies have
concentrated on cases where the effective Prandtl number is greater
than unity
\citep[e.g.][]{2016AdSpR..58.1475O,2017ApJ...851...74B,2018PhFl...30d6602K}. This
was motivated by the observation that the overall velocities are
decreased in high-$\Pra$ convection. However, this also coincides with
more effective downward flux of angular momentum, exacerbating the
problems related to anti-solar differential rotation
\cite{2018PhFl...30d6602K}. Another recent study
\cite{2023A&A...669A..98K} confirmed these ideas and showed that the
Prandtl number dependence is relevant in the regime $\Pra\gtrsim1$,
whereas for $\Pra\lesssim1$, the parameter regime relevant for the
Sun, no statistically significant dependence was detected.

Finally, the role of magnetism in shaping the solar rotation profile
is also a viable option to explain the convective conundrum. Whereas
early forays into this field yielded somewhat contradictory results
with some studies finding essentially no dependence on magnetic fields
\citep{KKKBOP15}, others reported a flip from anti-solar to solar-like
differential rotation \citep{FF14,2015ApJ...810...80S}. All of these
simulations were made at relatively modest magnetic Reynolds numbers,
and it is likely that a SSD was not excited in these models. The
recent high-resolution implicit large-eddy simulations (iLES)
\citep{2021NatAs...5.1100H,2022ApJ...933..199H}, have reached a regime
where small-scale magnetic fields are generated throughout the
convection zone and turn a hydrodynamically anti-solar run to a
solar-like solution at the highest resolution. Somewhat worryingly,
these simulations have yet to show convergence as a function of
resolution such that the flows at large scales changes significantly
even between the two largest resolutions. Recently,
\cite{2023A&A...669A..98K} reported that it is easier to excite
solar-like differential rotation for higher $\ReM$ from simulations
with explicit diffusivities where $\ReM$ exceeded the threshold for
SSD. However, this effect is much less drastic than in the iLES
simulations.

The radial shear in the solar convection zone occurs predominantly in
the boundary layers which are difficult to incorporate in global
simulations. The near-surface shear layer is thought to be generated
in the weakly rotationally constrained small-scale convection in the
outermost parts of the solar convection zone \citep[e.g.][]{Ki16} or
due to gyroscopic pumping effects
\citep[e.g.][]{2011ApJ...743...79M}. Capturing this in global
simulations is challenging because a very high resolution is required
to capture the near-surface small-scale convection resulting from a
steep decrease of fluid density. First such simulations were presented
by \cite{HRY15a}, who were able to capture some aspects of the NSSL.
However, these simulations were hydrodynamic and no corresponding
dynamo solutions have been presented so far. The NSSL has also been
suggested to shape the global solar dynamo \cite{Br05}, but no direct
evidence supporting or refuting this theory is currently available.

The other boundary layer at the interface of the convective and
radiative layers, the tachocline, is perhaps even more challenging to
capture in simulations. The main challenge is that the solar
tachocline is very thin (certainly less than five per cent of solar
radius, but likely much less), and it has been confined now for five
billion years. Estimates of radiative spreading for the Sun suggest
that the tachocline should be much thicker at the current age of the
Sun so there has to be a mechanims preventing this. Several magnetic
scenarios have been invoked to explain this, including a dipolar
fossil field in the radiative core or a cyclic dynamo in the
convection zone. Some current simulations do exhibit tachocline-like
features, but they are spreading into the radiative core at rates that
are much higher than expected for the Sun
\citep[e.g.][]{BMT11,2017ApJ...836..192B}. Furthermore, iLES models
also produce tachoclines at relatively low resolutions, although their
confinement mechanism is yet to be understood
\citep{GSKM13,GSdGDPKM15,Guerrero+22}. In all of the aforementioned
simulations the diffusivities in the radiative interior were either
explicitly or implicitly greatly reduced. In apparent contradiction to
these models, \cite{2022ApJ...940L..50M} were able to obtain a
relatively thin tachocline and essentially rigidly rotating radiative
core in a simulation where the diffusivities were not decreased but
which housed a cycling non-axissymmetric dynamo in the convection
zone. Furthermore, this model has also strong horizontal flows in the
radiative interior. The actual process of tachocline confinement is
still unclear in this case, although it does share some
characteristics with the cyclic dynamo confinement process suggested
by \cite{2001SoPh..203..195F}; see also \cite{2017A&A...601A..47B}.

\begin{figure*}[t]
\begin{centering}
  \includegraphics[width=0.75\textwidth]{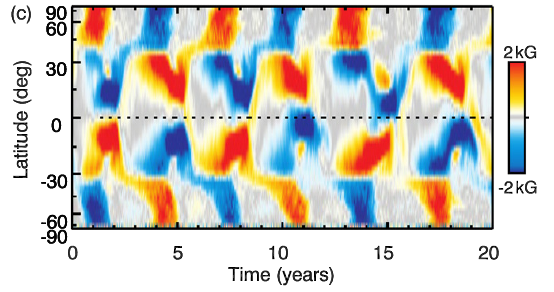}
  \includegraphics[width=0.75\textwidth]{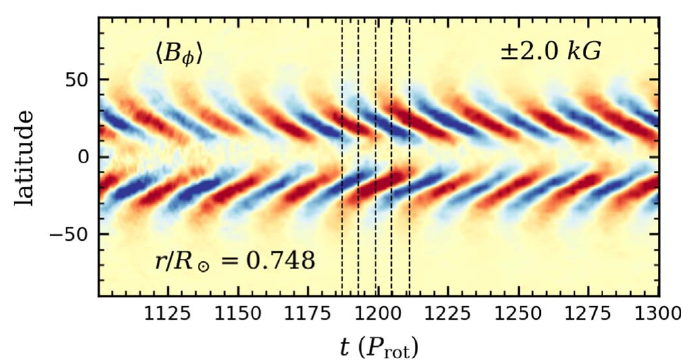}
  \includegraphics[width=0.75\textwidth]{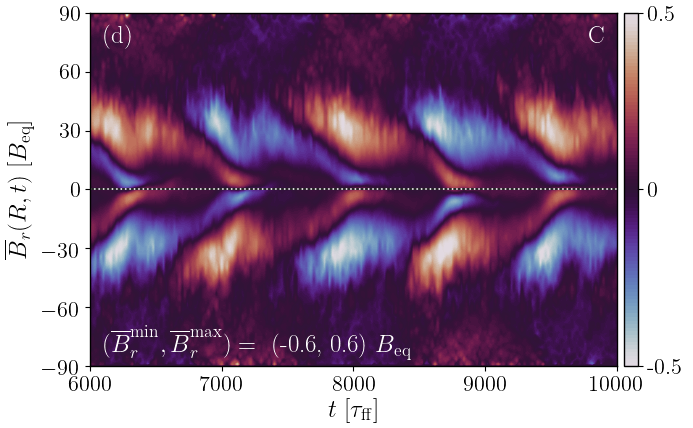}
\caption{Butterfly diagrams from \cite{ABMT15} (top),
  \cite{2020ApJ...892..106M} (middle; \textcopyright AAS. Reproduced with permission) and \cite{2022ApJ...931L..17K}
  (bottom).}
\label{fig:bfly}
\end{centering}
\end{figure*}

Given the difficulties in reproducing the solar flows, it is then
hardly surprising that dynamo simulations have had a hard time
reproducing the solar large-scale magnetism. The most severe issue is
the difficulty in obtaining solar-like equatorward migration of
activity belts in simulations with solar-like differential
rotation. Nevertheless, several simulations have appeared showing
equatorward migration and which capture many aspects of solar
observations. For example, \cite{KMB12} reported equatorward migration
from spherical wedge simulations that were later shown to be in
accordance with a Parker--Yoshimura dynamo wave resulting from a
mid-latitude minimum of angular velocity which is not present in the
Sun \citep{WKKB14,2018A&A...609A..51W}. A similar mid-latitude dip is
seen also in the equatorward migrating solutions of
\cite{ABMT15}. Further examples of equatorward propagating solutions
have been reported by \cite{DWBG16}, \cite{2020ApJ...892..106M},
\cite{SBCBN17,2018ApJ...863...35S}, and \cite{2022ApJ...926...21B};
see also \Fig{fig:bfly}. The latter authors argued that a non-linear
interplay between the magnetic fields and differential rotation can
lead to solar-like long period cyclic dynamos.
Another recent example shows equatorward migration near the surface
but poleward migration at depth in a star-in-a-box model
\citep{2022ApJ...931L..17K}, where a spherical star is embedded into a
Cartesian cube. In such models the boundary of the star is immersed
into the domain and, in theory, allows for a more realistic magnetic
boundary condition. This was shown to be important in that if the
exterior was made a poor conductor, the global dynamo solution changed
from oscillatory to quasi-static. This confirms earlier results of
\citep{WKMB13,WKKB16}, where the influence of a simplified coronal
layer as upper boundary on the flow and magnetic field evolution was
studied.

These simulation results and their interpretation add to the ongoing
debate regarding the location and dominant physical mechanisms of the
solar global dynamo. The current state of affairs is particularly
clearly manifested by the wide variety of mean-field models that have
been put forward to explain the solar cycle. A popular class of models
include flux-transport \citep[e.g.][]{DC99} and Babcock-Leighton
\citep[e.g.][]{2017A&A...599A..52C} dynamos where a minimal set of
physically plausible ingredients, such as differential rotation and
meridional circulation and the decay of active regions near the solar
surface are taken into account. These models typically rely on
buoyantly rising flux ropes \citep[e.g.][]{1995ApJ...441..886C} that
are the result of strong shear in the tachocline, and the turbulent
flows in the convection zone play only the role of turbulent
diffusion. On the other hand, distributed turbulent dynamos take into
acount a variety of effects that arise in mean-field electrodynamics
and assume magnetic field generation throughout the convection zone
\citep[e.g.][]{BMT92,KKT06,2013ApJ...776...36P}. The obvious drawback
of mean-field models is that the individual effects can be adjusted
which leads to a great temptation to finetune the models. The main
advantage of 3D simulations is that this freedom is greatly reduced
(although by no means completely eliminated). A major part of the
debate regarding the solar dynamo revolves around the relevance of the
tachocline.

Unfortunately the advent of 3D simulations with and without
tachoclines has not provided conclusive evidence one way or the other.
For example, \cite{GSdGDPKM15} studied both cases with {\sc Eulag}
simulations and found that dynamos operating solely in the convection
zone had shorter cycles and intermittent turns-off of activity.  On
the other hand, dynamos operating at tachocline levels result in long,
coherent, magnetic cycles. Whereas the dynamos operating in the
convection zone are understood as distributed $\alpha\Omega$ dynamos,
those operating at the tachocline may be of $\alpha^2\Omega$ type,
with the $\alpha$ effect generated by instabilities that extract
energy from the magnetic field \citep[e.g., Tayler or buoyancy
  instabilities; see][]{2019ApJ...880....6G}. Furthermore,
simulations with \citep[][]{2022ApJ...931L..17K} and without
\citep[][]{KMB12,2018A&A...616A..72W} tachoclines from other models
produce cyclic dynamos that share some characteristics of the solar
cycle. A general conclusion is that long cycles appear to be generated
at depth while shorter ones have their origin near the surface
\citep[e.g.][]{KKOBWKP16}.

Apart from the non-linear dynamo invoked by \cite{SBCBN17} and
\cite{2022ApJ...926...21B}, another physical mechanism proposed to be
responsible for the equatorward migration include helicity inversion
in the deep parts of the convection zone. This is typically
encountered in overshoot regions below the convection zone
\citep[e.g.][]{OSB01}. However, a much deeper helicity inversion was
obtained in simulations by \cite{DWBG16}, and which resulted in a
change of the propagation direction of the dynamo wave in accordance
with the Parker-Yoshimura rule. In these simulations convection was,
however, inefficient in the bulk of the convection zone which is not
the situation in the solar convection zone. Such reversed helicity
configuration can perhaps arise if much of the convection zone is
stably stratified as in the proposed entropy rain scenario, but there
is currently no simulation that has produced this.

A yet further possibility is that the solar dynamo is driven
predominantly by the kinetic helicity as in classical $\alpha^2$
dynamos that can lead to equatorward migration if the $\alpha$ effect
has a sign change at the equator \citep[][]{BS87,Ra87}. Such helicity
profile is expected due to symmetry arguments theoretically, and is a
standard outcome in convection simulations \citep[e.g.][]{BMT04}. The
equatorward migrating dynamo wave has been demonstrated by
three-dimensional forced turbulence simulations
\citep[e.g.][]{MTKB10,WBM11}, but no definitive evidence from
convection exists.

There have also been attempts to capture the long term modulations in
dynamo cycles from simulations. These studies have extended the
simulations to cover several tens of cycles corresponding to up to a
millenium in solar time \cite[e.g.][]{PC14,ABMT15,KKOBWKP16}. These
simulations revealed modulation of activity and occasional periods of
low activity reminiscent of grand minima
\citep{ABMT15,KKOBWKP16}. Such grand minima states can arise due to an
interplay of symmetric and anti-symmetric dynamo modes
\citep[e.g.][]{1997A&A...322.1007T} or stochastic fluctuations in the
buoyancy driving or in the conventional $\alpha$ effect
\citep[e.g.][]{2000A&A...359..364O,2008AN....329..351B}. However, the
modulations and minima in simulations are clearly weaker compared to
the solar observations. This is perhaps not too surprising because to
run these simulations sufficiently long they need to be done at modest
resolutions and cannot therefore be highly supercritical.

\subsection{The Sun at different ages}
\label{sec:sunintime}

During its 4.5 billion years of evolution, the Sun has experienced
various changes in its internal constitution, and therefore, the
extent of its convection zones and resulting large-scale flows and
magnetic fields. In this section we describe numerical simulations of
the Sun, or Sun-like stars, corresponding to these evolutionary stages
from the formation to the current age.

\subsubsection{The pre-main sequence phase}
\label{sec:ttauri}

Significant structural changes occurred early in the solar evolution
during the pre-main sequence (PMS) stage, as the newly formed object
is still contracting. Objects at this stage, with masses similar to
the solar mass, are called TTauri stars. While the temperature at the
center of the protostar is still increasing, the opacity of the gas is
high and the transport of energy occurs entirely due to
convection. The actual rotation rate of the Sun in the TTauri phase is
unknown, but models can be constructed to characterise it
\citep[e.g.][]{2020A&A...635A.170A}. Moreover, observations of open
clusters have found distributions of rotational periods between
roughly $1$ and $10$ days for solar-like stars with ages around $3$
Myrs \citep[see e.g.,][]{Gallet+13}.

The large-scale magnetic fields of TTauri stars are predominantly
dipolar with field strengths of the order of kG
\citep[e.g.][]{2007ApJ...664..975J}. Fields with a similar topology
are also often observed in low mass, fully convective and rapidly
rotating M dwarfs \citep[e.g.][]{2021A&ARv..29....1K}. Therefore,
despite the difference in mass, simulations of TTauri stars and
low-mass M dwarfs are, to some extent, comparable.  However, the
latter will be discussed in detail in \Sec{sec:fullco}. Following the
evolution further, the protoplanetary disc disappears after about
$10^6-10^7$ years since the beginning of the collapse. The protostar
continues to contract, and therefore its angular velocity
increases. Simultaneously, the star starts to develop a radiative
core. Both, the angular velocity and the radiative zone increase
before the star reaches the zero age main sequence (ZAMS) after about
$5\times 10^7$ years. As mentioned above, during the TTauri phase the
magnetic field of a solar-like star is mainly dipolar. Observations
suggest increasing complexity of the magnetic topology once the star
develops a radiative zone \citep{Gregory+12}.

There are currently only a few simulation studies that specifically
target dynamos in the TTauri and PMS phases of stellar evolution. One
such example is the study of \cite{Zaire+16} who considered models in
the fully and partially convective phases of PMS evolution. While the
differential rotation was more pronounced in the latter evolutionary
phase, the resulting quasi-steady predominantly quadrupolar magnetic
field configurations were quite similar. A more complete study was
performed by \cite{Emeriau+17}, where five epochs between the TTauri
stage and the ZAMS were studied for a $1 M_{\odot}$ star. Each epoch
is characterised by a diffrent internal structure and rotation
period. The sequence of simulations shows a decreasing dipole
contribution to the magnetic field as a function of age. However, even
in the early fully convective phase, the dipole constitutes only about
ten per cent of the total magnetic energy. In this case the
azimuthally averaged large-scale magnetic field is cyclic with with
poleward migration, reminiscent of the simulations of fully convective
M dwarfs \citep[e.g.][]{2020ApJ...902L...3B,2021A&A...651A..66K}. The
transition between dipole-dominated and multipolar dynamos is
discussed in more detail from the perspective of simulations in
\Sec{sec:fullco}.

\subsubsection{Main sequence Sun-like stars: rotational evolution of differential rotation and dynamos}
\label{sec:solarrotevo}

On the main sequence, the rotation rate of stars decreases following
the observational \cite{1972ApJ...171..565S} law, associated to the
loss of angular momentum due to magnetized stellar winds.  There have
been some speculative ideas about the origin of the non-saturated and
saturated regimes of the rotation-activity relationship of stars
\citep[e.g.][]{1988ApJ...333..236K,2015ApJ...799L..23M}. For instance,
\cite{2011ApJ...743...48W} suggested that a turbulent (interface)
dynamo is at work in rapidly (slowly) rotating stars. However, the
fact that fully convective stars also follow the power law for slow
rotation \citep{2016Natur.535..526W} suggests the possibility of a
general dynamo theory for all main sequence stars. Nevertheless, to
the date, there is no agreement about this theory. One of the main
hindrances is the difficulty in observing differential rotation as a
function of $\Ro$
\cite[e.g.][]{2015A&A...576A..15R,2018Sci...361.1231B}, but a new
approach has been recently proposed by \cite{2022A&A...667A..50N}
using {\sc Kepler} data.  Equally difficult is obtaining unambiguous
measurements of dynamo cycle periods and systematics as a function of
rotation as already discussed.  Numerical simulations, on the other
hand, can be performed at arbitrary rotation rates corresponding to
different ages of the star. Below, we summarize the relevant findings
for the differential rotation and dynamos for a solar mass star from
its youth to the present age and beyond.

\begin{figure*}[t]
\begin{centering}
\includegraphics[width=0.47\textwidth]{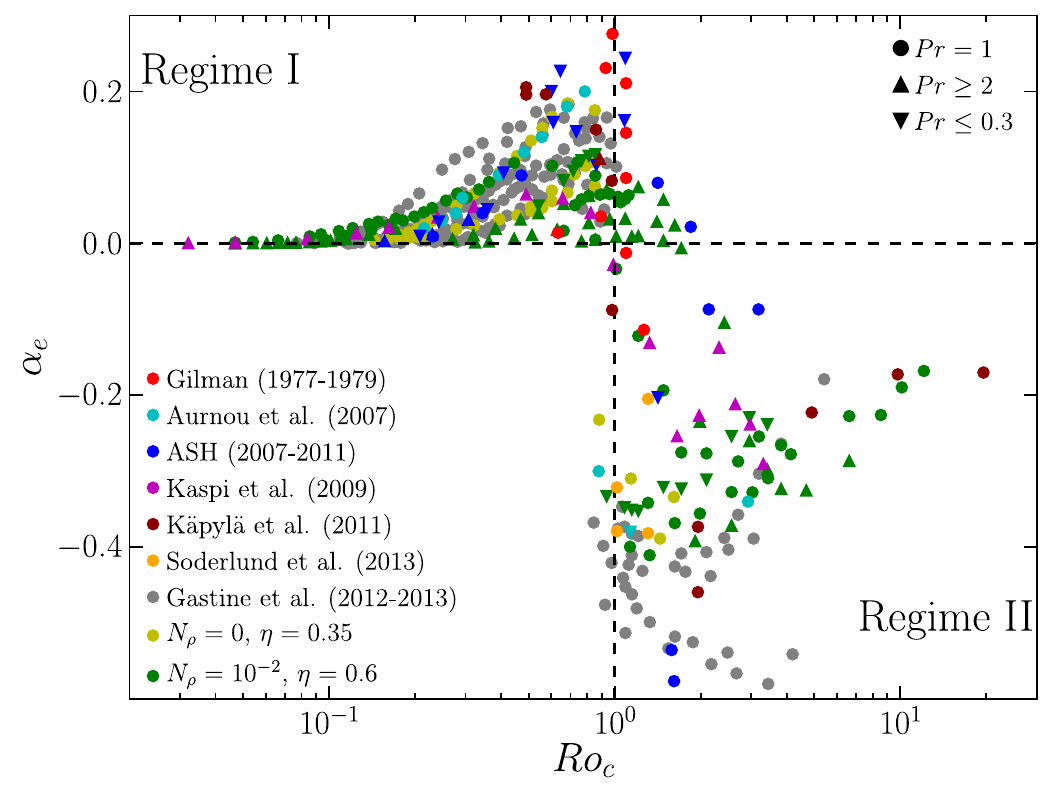}\includegraphics[width=0.53\textwidth]{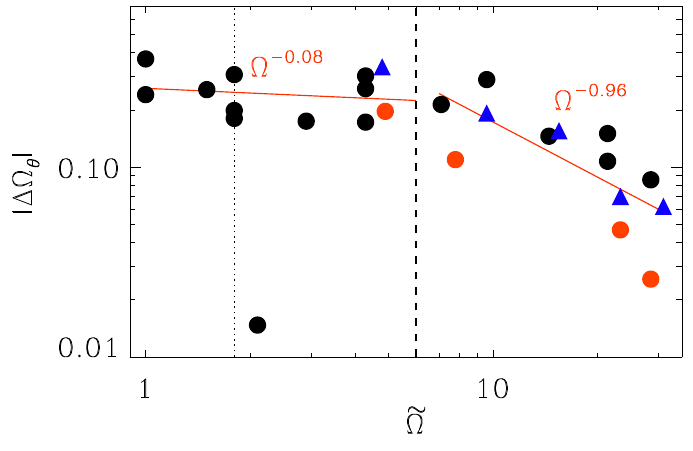}
\caption{Left: Measure of the radial differential rotation at the
  equator for several studies in the literature. Adapted from
  \cite{GYMRW14}. Right: Modulus of the absolute latitudinal
  differential rotation from simulations of a solar-like star
  \cite{2018A&A...616A.160V}, where $\tilde\Omega$ is the rotation
  rate normalized by the solar rotation (\textcopyright
  ESO. Reproduced with permission). The dotted line corresponds to the
  transition of anti-solar to solar-like differential rotation at
  $\Ro\approx0.35$ and the dashed line separates the two regimes of
  differential rotation dependence at $\Ro\approx0.08$.}
\label{fig:DR}
\end{centering}
\end{figure*}

The Rossby number dependence of large-scale mean flows has been
studied in various papers
\citep[e.g.][]{2007ApJ...669.1190B,KMGBC11,KMB11,GSKM13,GYMRW14,FM15,2017ApJ...836..192B}. These
studies considered different rotation rates for roughly the same
structural model resembling the solar interior. Irrespective of the
numerical scheme, the results confirmed that the relative radial
differential rotation $\Delta \Omega/\Omega$ changes from positive
(solar-like differential rotation), for small $\Ro$, to negative
(anti-solar) for large $\Ro$ \citep[see left panel of \Fig{fig:DR},
  adapted from ][]{GYMRW14}, with the transition happening near
$\Ro=1$. In \cite{2018A&A...616A.160V}, the modulus of the absolute
differential rotation was also found to decrease rapidly with the
rotation rate for $\Ro \lesssim 0.1$; see right panel of \Fig{fig:DR}
and Figure 8 of \cite{2022ApJ...926...21B}.  This decrease, however,
can be due to low supercriticality of convection at such low $\Ro$.
General consensus from simulations is that for sufficiently rapid
rotation the differential rotation is negligibly small. This has
implications on the theoretical interpretation of dynamos in the
classical mean-field dynamo framework; see \Sec{sec:conMF}. Another
characteristic is the appearance of non-axisymmetric convective modes,
or active nests, in the rapidly rotating regime, $\Ro \ll 1$
\citep{BBBMT08}. Such non-axisymmetric convection has recently been
suggested to be the origin of stellar active longitudes
\citep{2022ApJ...928...51B}. Regarding the structure of convection, it
is evident from all simulations that rotation breaks the broad
convective cells observed in non-rotating or slowly rotating
simulations. Quantitatively, \cite{FH16b} found that the harmonic
degree where the spectrum has a maximum, $\ell_{\rm peak}$, scales
with the Rossby number as $\ell_{\rm peak}\sim \Ro^{-1/2}$ \citep[see
  also][]{2018A&A...616A.160V}. This means that the faster the
rotation, the smaller the scales where most of the kinetic energy is
contained.  This applies to regions of the convection zone where
$\Ro\lesssim1$; in the near-surface layers $\Ro\gg1$ even in the most
rapidly rotating stars, and the size of photospheric convection cells
is likely independent of stellar rotation.

As discussed in \Sec{sec:observations}, the rotation of stars slows
down due to magnetic braking as they age. The young Sun was therefore
a much faster rotator than what it is today. Simulations of rapidly
rotating ($\Ro\lesssim0.1$) young solar-like stars have been performed
by several groups using various numerical methods. Quite surprisingly,
the results in this parameter regime are rather inhomogeneous: there
are three distinct dynamo modes that have been reported from such
studies. First, there is the large-scale dipole-dominated solutions
\citep[e.g.][]{GDW12,YGCR15,2022MNRAS.517.3392Z} that are reminiscent
of the geodynamo and those of Saturn and Jupiter. Another outcome is
that the large-scale magnetic fields are dominated by a
non-axisymmetric $m=1$ mode that often propagates either in retro- or
prograde fashion
\citep[e.g.][]{CKMB14,YGCR15,2018A&A...616A.160V,2021A&A...645A.141V,2022A&A...667A.164N}.
Finally, some simulations produce predominantly axisymmetric but
multipolar large-scale fields in such rapidly rotating setting
\citep[e.g.][]{2018ApJ...863...35S}. It is unclear why there is such a
variety in magnetic field topologies in this regime. A possible cause
is that the dynamo is sensitive to relatively minor differences in the
boundary conditions \citep[see, e.g.][]{WKKB16} and/or other details
of the simulation setups \citep{2021MNRAS.507L..67O}.

\begin{figure*}[t!]
\begin{centering}
 \includegraphics[width=0.45\textwidth]{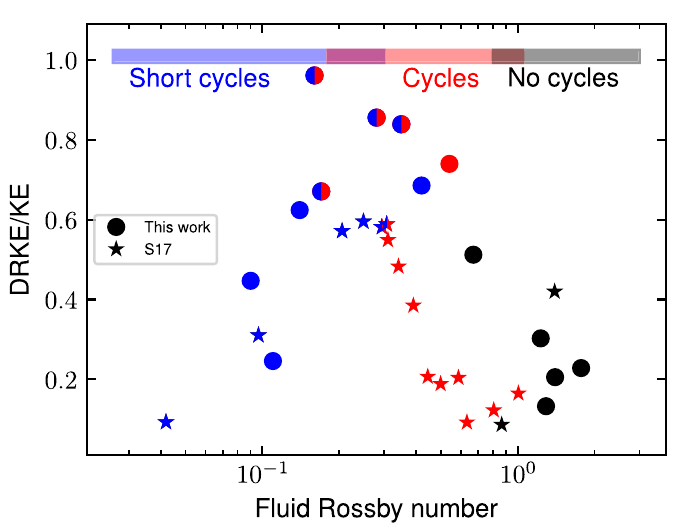}\includegraphics[width=0.55\textwidth]{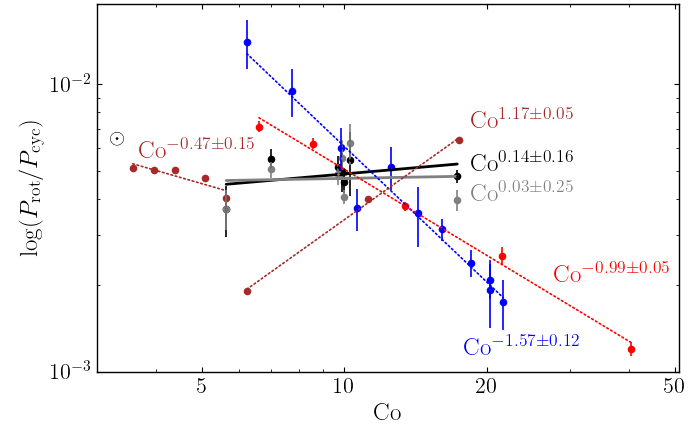}
 \caption{Left: Summary of the fraction of differential rotation
   energy from total kinetic energy as a function of $\Ro$ from the
   simulations of \cite{SBCBN17} and \cite{2022ApJ...926...21B}. The
   colours of the symbols indicate the type, or the lack of, magnetic
   cycles. Adapted from \cite{2022ApJ...926...21B}. Right: Comparison
   of the ratio of the rotation period and the cycle periods as a
   function of the Coriolis number ($\propto \Ro^{-1}$) from the
   studies of \cite{2018ApJ...863...35S} (blue),
   \cite{2018A&A...616A..72W} (red), \cite{2019ApJ...880....6G}, and
   \cite{2022ApJ...931L..17K}. Estimated location of the Sun is
   indicated by the symbol $\odot$. Adapted from
   \cite{2022ApJ...931L..17K}.}
\label{fig:protpcyc}
\end{centering}
\end{figure*}

Two further regimes of dynamos can be distinguished on either side of
transition between solar-like to anti-solar differential rotation.
When rotation is rapid enough such that a solar-like differential
rotation is produced, the large-scale fields are predominantly
axisymmetric and often cyclic
\citep[e.g.][]{BBBMT10,GCS10,KMB12,NBBMT13,ABMT15,2018A&A...616A..72W,2020ApJ...892..106M}.
As mentioned above, in some cases this behavior continues to much more
rapid rotation whereas in others non-axisymmetric or dipolar dynamos
modes take over. When rotation is slow enough and the differential
rotation is anti-solar, the magnetic fields are predominantly
axisymmetric and quasi-steady
\citep[e.g.][]{2017A&A...599A...4K,2018ApJ...863...35S,2018A&A...616A..72W}.
In the transition between the two regimes, the dynamo may excite the
two modes simultaneously \citep{2019ApJ...886...21V}.  The appearance
of cycles appears to be related to the strength of the differential
rotation such that long decadal cycles, such as in the Sun, appear in
a relatively narrow range of Rossby numbers where the differential
rotation is strong \citep[][see the left panel of
  \Fig{fig:protpcyc}]{2020A&A...642A..66W,2022ApJ...926...21B}.

Regardless of the commonalities between different modeling approaches,
the simulated magnetic cycle seem to be sensitive to the subtleties of
the models, and there is no clear agreement regarding the scaling of
cycle periods as a function of rotation as can be seen in the right
panel of \Fig{fig:protpcyc} compiling the results of
\cite{2018ApJ...863...35S}, \cite{2018A&A...616A..72W},
\cite{2019ApJ...880....6G}, and
\cite{2022ApJ...931L..17K}. Furthermore, none of the scaling laws from
simulations seem to unambiguously agree with the observed cycles that
are sometime grouped in activity branches with $\Prot/\Pcyc\propto
\Co^\alpha$, where $\alpha > 0$
\citep[e.g.][]{SB99,2017ApJ...845...79B}. It is worth mentioning,
however, that the observational results may also present problems and,
depending on the used analysis techniques, the distinct activity
branches may not even exist \citep[e.g.][]{Bonanno+22}.

As shown in \Fig{fig:DR} (left), the amplitude of the differential
rotation has maxima on either side of the solar-like to anti-solar
transition. The strong shear in the anti-solar regime has been
speculated to lead to enhanced magnetic activity
\citep{2018ApJ...855L..22B}. While this has not been systematically
studied, some evidence of enhanced magnetic energy for anti-solar
differential rotation has been found in simulations
\citep[e.g.][]{KKKBOP15,2020A&A...642A..66W,2022ApJ...926...21B}. A
much more unambiguous observational fact is that the stellar magnetic
activity and magnetic fields saturate for $\Ro\lesssim0.1$
\citep[e.g.][see also
  \Sec{sec:observations}]{2022A&A...662A..41R}. This does not appear
to happen in simulations, where the magnetic energy typically
increases with rotation even for the fastest rotation considered so
far \citep[e.g.][]{2020A&A...642A..66W}. Furthermore, the ratio of
magnetic to kinetic energy increases roughly proportional to
$\Ro^{-1}$ \citep{2019ApJ...876...83A,2020A&A...642A..66W} in
accordance with Magneto-Archimedean-Coriolis (MAC)
balance. \cite{2022ApJ...926...21B} showed that the large-scale
surface fields follow even a steeper increasing trend $\propto
\Ro^{-1.4}$ with decreasing Rossby number. However, this is consistent
with stellar observation in the magnetically non-saturated regime
\citep[][]{2019ApJ...886..120S,2022ApJ...926...21B}.  Nevertheless, it
is unclear why the simulations deviate from observations in that they
do not show indications of saturation when the Rossby number is
decreased. It is plausible that missing surface physics, such as the
lack of spot formation in the simulations contributes to this issue.

\subsection{Stars other than the Sun}

Most stars are not like the Sun. A large majority (in our galaxy,
anyway) are M dwarfs, which are less massive than the Sun,
considerably less luminous (ranging down to about $10^{-3}$
$L_{\odot}$), and in some cases convective throughout their
interiors. At the other end of the H-R diagram, stars more massive
than the Sun have convective cores and predominantly stable
(radiative) envelopes, accompanied in some cases by thin near-surface
convection zones. These high-mass stars can be thousands of times more
luminous than the Sun; thus across the main sequence, luminosities
vary by a factor of more than a million. These enormous variations in
luminosity, and in the geometry and stratification as well, must
influence the convection and dynamo action occurring in these stars.
In this section, we therefore briefly describe our current
understanding of dynamo action in main-sequence stars at lower and
higher masses. Our focus here is on what has been revealed by basic
theory and by simulations.

From the standpoint of dynamo theory, stars vary not just in their
luminosity but also their rotation rate, their geometry, their
stratification, and their microphysics. For example the core of a
massive star is only weakly stratified, whereas an M dwarf (or the
convective envelope of a Sun-like stars) is much more strongly
stratified.  The microscopic diffusivities vary in such a way that
$\PrM$ is very small in many stellar convection zones but greater than
unity in some
\citep[e.g.][]{2019ApJ...876...83A,2022ApJS..262...19J}. Many of these
variations are intertwined. For example, the influence rotation has on
the convection is typically encapsulated by some version of the Rossby
number $\Ro \sim \Prot/\tauconv$; because $\tauconv$ depends on the
flow velocity, the rotational influence can vary from star to star
(and with radius within a given star) even if the rotation period is
constant.

Faced with the bewildering array of possible variations in all of
these parameters, would-be simulators tend to try one of two different
approaches. One is to focus on a particular physical effect --
stratification, for example -- and try to elucidate how this affects
convection, differential rotation, and magnetism, typically in an
idealized setting (e.g., a Cartesian layer). Another is to attempt to
model a given astrophysical object in some detail, adopting (e.g.)
spherical coordinate systems and stratifications (of density,
temperature, or entropy) that mimic those in a 1D stellar model. In
neither approach is it possible for simulations to approach the actual
parameter regimes attained in real stars, although some balances can
be recovered that help in understanding, e.g.\ the type of
differential rotation that occur in real stars. Before diving into a
``by object'' discussion, we turn first to a high-level overview of
the effects of rotation, stratification, and geometry on dynamo action
in general. Then we turn to summaries of how these effects play out in
specific types of stars.

\subsubsection{Effects of rotation, stratification, and geometry in stellar models}

All stars rotate, all have some level of density stratification, and
all convect in a spherical geometry. Here we provide a very brief
overview of how each of these effects likely influence the flows and
magnetism. Perhaps the clearest message from the last few decades of
research on convection and dynamo action is, ``rotation matters a
lot.'' It affects the convective flows, it affects their transport of
heat and angular momentum, and it affects the magnetism these flows
can build. Some of these effects (e.g., how rotation affects magnetic
field morphology) are understood only qualitatively, whereas for
others (e.g., how it affects heat transport) there is now a more
quantitative picture.

To begin, consider the influence of rotation on the convective flows
in the absence of magnetism. It is well-known that rotation tends to
stabilise a system against convection, increasing the critical
Rayleigh number for onset ($\Ra_{\rm c} \propto \Ek^{-4/3}$, with
$\Ek$ the Ekman number defined previously), while the most unstable
wavenumber shifts to higher wavenumber ($k$) with more rapid rotation
rate (lower $\Ek$) \citep[e.g.][]{Ch61}. (Less obviously, consider a
numerical simulation not too far from onset -- more specifically, one
for which ``diffusion free'' scalings do not yet apply -- with, for
example, a fixed heat flux or temperature contrast, and also some
fixed rotation rate. In this system changing the numerical
diffusivities -- or in an implicit LES simulation changing the
resolution -- will also affect the rotational influence, via changing
$\Ek$, and so also the flows.)  In a global spherical geometry,
whenever rotation plays a major role in the dynamics the most
prominent convective modes close to onset are wave-like convective
rolls, aligned with the rotation axis and arising from the
conservation of potential vorticity, variously called ``thermal Rossby
waves,'' ``Busse columns,'' or ``banana cells'' depending on the
community \citep[see,
  e.g.][]{1970JFM....44..441B,Gi77,1983PhyD....9..287B,FH16b,2022A&A...666A.135B}. The
systematic prograde tilt of these cells, and the associated Reynolds
stress, plays a major role in redistributing angular momentum in most
global-scale simulations of stellar convection; the differential
rotation is then also a strong function of rotation rate (Rossby
number), as summarized elsewhere \citep[e.g.][]{GYMRW14}.

The heat transport by the convection is also strongly influenced by
rotation: for a given fixed heat flux, the convection tends to become
``less efficient'' as the rotation rate increases -- that is, it
requires a larger entropy gradient to carry the same flux \citep[see,
  e.g., discussions
  in][]{1979GApFD..12..139S,2020PhRvR...2d3115A}. Quantitatively,
several theoretical approaches supported by numerical simulations --
including appeals to a balance between Coriolis, inertial, and
buoyancy forces (so-called ``CIA balance'')
\citep[e.g][]{2021PNAS..11822518V}, ``rotating mixing length'' theory
\citep{1979GApFD..12..139S,2014ApJ...791...13B,2020MNRAS.493.5233C}
asymptotic theory at low Rossby number \citep{2012GApFD.106..392J} --
yield the same diffusion-free dependence of key quantities, like the
temperature gradient, when rotation is rapid enough. In these models
the amplitude of the convective motions goes down at higher rotation
rates, the temperature gradient goes up, and the typical wavenumber of
the flow increases, with important consequences for the dynamo.

The dependence of dynamo action itself on rotation is less
quantitatively understood, though it is still possible to make some
broad statements that are supported by theory and simulation. When
rotation is dynamically significant, it is widely thought to affect
the strength and morphology of dynamo-generated magnetic fields as
discussed in more detail in \Sec{sec:solarrotevo}.  \citep[see also,
  e.g.,
][]{2013MNRAS.431L..78S,2019ApJ...876...83A,2021MNRAS.507L..67O}.
Furthermore, because the envelope convection zones of low-mass stars
are strongly stratified (in density and temperature), many authors
have also sought to understand what role this stratification plays in
determining various properties of the flows and magnetism. We mention
here only a few.  Broadly, the presence of density stratification
breaks the up-down symmetry in Boussinesq systems, so it changes the
convective dynamics -- and also presumably the magnetism -- in a
variety of ways. Strongly stratified systems tend to have strong,
narrow downflows and broader, weaker upflows
\citep[e.g.][]{HTM84,CBTMH91}; their energy dissipation budget is very
different \citep[e.g.,][]{1975JFM....68..721H,2017ApJ...845L..17C};
they can establish different profiles of kinetic helicty
\citep{DWBG16} and may drive different types of zonal flows
\citep[e.g.,][]{2009GApFD.103...31G}. If diffusion-free scalings (like
mixing-length theory) apply, or simply on dimensional grounds, the
velocity of the convective flows is expected to vary in amplitude
across a stratified convection zone, in a way not found in Boussinesq
systems. This basic fact has profound consequences for the convective
dynamics in a star like the Sun, since we then expect regions near the
photosphere (which are very low-density) to undergo such rapid
convective motions that their Rossby number is enormous (i.e., the
influence of rotation is negligible there); meanwhile the deeper flows
must be influenced by rotation, at least within some finite distance
of the transition to the radiative interior. The presence of density
stratification may make it harder to sustain a strong global-scale
dipole in some regimes \citep[e.g.,][]{GDW12}, but it is also clear
that highly ordered fields are still realizable even when the
stratification is strong (see discussions in
\citealt{2015MNRAS.448.2055R, 2020PEPI..30706542M,
  2022MNRAS.517.3392Z}).

Relatively little work has directly addressed the influence of the
\emph{geometry} of the convection zone, while keeping other factors
constant. We briefly note only a few specific results. \citet{GD08}
showed that, in rotating Boussinesq MHD simulations with fixed
temperature boundaries, changing the aspect ratio of the domain (i.e.,
changing the depth of the convection zone) had a strong effect on the
nature of their dynamo solutions. In deep domains, they found steady
``Earth-like'' dipolar solutions; if the convection zone was gradually
made thinner, they found a transition to ``Sun-like'' dynamo wave
solutions.  Separately, \citet{2022ApJ...938...65C} have recently
investigated the role of geometry in determining the Reynolds stresses
-- and hence the differential rotation -- in anelastic simulations of
rotating convective envelopes. They argue that the well-known
transition from solar to anti-solar differential rotation occurred
when columnar convective structures attained a diameter roughly
equivalent to the shell depth. In the sections that follow, we
explore how these different dynamical trends play out in models of
specific stars.

\subsubsection{Dynamo action in high-mass stars: convective cores and
  radiative envelopes}

Massive stars on the main sequence possess convective cores, with a
predominantly stable radiative envelope above this. The high
luminosities established by nuclear fusion in these stars mean that we
expect the convection to be vigorous; the accompanying high
temperatures mean that we expect these convection motions should
generally act as magnetic dynamos, since all plausible estimates of
the magnetic Reynolds number $\ReM$ are very high; see
\Table{tab:dimpar}. It may well be that dynamo action is occurring in
the radiative layers as well, as discussed below. Both the flows and
fields have lately been targets of intense scrutiny -- partly because
asteroseismology has begun to provide powerful new constraints on this
topic, particularly for evolved stars (see, e.g.,
\citealt{2016Natur.529..364S, 2023MNRAS.521.5372J}) and also because
of the implications these hold for later stages of stellar evolution
\citep[e.g.,][]{2019ApJ...881L...1F}. In this section, we briefly
review what has been learned by simulations focusing on these types of
stars. A more thorough description can be found in
\citet{2017LRSP...14....4B}.

Early global simulations covering some aspects of the problem include
\citet{2003ASPC..293..147K}, \citet{browning_etal2004}, and
\citet{brun_et_al_2005}.  Taken together, these papers provided the
first numerical estimates of the flows and magnetism that might be
generated in 3D massive star cores. They also gave some estimates of
the overshooting and penetration from the convective zone into the
surrounding stable envelope, the gravity wave response there, and the
differential rotation arising from the interplay of convection,
magnetism, and rotation. Later work has pushed towards more realistic
(turbulent) flows, has examined a variety of initial states (e.g.,
strong magnetic ``fossil'' fields) and srutinized each facet of this
complicated problem more systematically. We refer the interested
reader to \citet{2007ApJ...665..690M}, \citet{2009ApJ...705.1000F},
\citet{2013ApJ...773..137G}, \citet{2015ApJ...815L..30R},
\citet{2016ApJ...829...92A}, \citet{2019ApJ...876....4E},
\citet{2022A&A...667A..43B}, and \citet{2023MNRAS.519.5333B} as a
representative sample of relevant work.

There has lately also been sustained interest in the topic of
magnetism generated by dynamo action in the radiative envelope,
typically as a result of the interaction between shear and magnetic
instabilities \citep[e.g.,][]{Spruit:1999vt}. Numerical work --
beginning with \citet{2006A&A...449..451B} and followed by many others
since (e.g., \citealt{2007A&A...474..145Z},
\citealt{2010ApJ...724L..34D}, \citealt{2015A&A...575A.106J},
\citealt{2023Sci...379..300P}, \citealt{2023MNRAS.521.5372J}) -- has
examined the circumstances under which such dynamo action could occur,
and the strength of the resulting fields.

Taken as a whole, these simulations are unequivocal about a few
points. Very strong fields can plausibly be established by dynamo
action in the cores of some massive stars, whether rotation is
dynamically significant or not; for example, because the convection is
rapid, even the ``equipartition-scale'' field (equating $\rho \uconv^2$
with $B^2/(4\pi)$ and assuming MLT scalings for the convective
velocity $\uconv$) would suggest $B \ge 10^6$ G in the interior of a
B-type star. Some of these stars rotate very rapidly, and the
saturation strength of the field in this case is less clear (see,
e.g., discussions in \citealt{2019ApJ...876...83A}), but it seems
likely that strong fields are the norm rather than the exception. The
differential rotation established within the core is less certain,
because it is surely influenced by the strength of the magnetism --
which is likewise influenced by the shear -- but broadly these
simulations appear to obey trends similar to those in simulations of
convective envelopes.  ``Solar-like'' differential rotation (i.e.,
with a prograde equator) is established when rotation is dynamically
significant and the magnetism is not too strong; ``anti-solar''
profiles arise when the influence of rotation is weaker; magnetism
reduces the shear and may yield solid-body rotation if it grows strong
enough. Convective motions overshoot into the radiative envelope,
though the extent of this effect is still being actively investigated
(e.g., \citealt{2022ApJ...926..169A}, \citealt{2023MNRAS.519.5333B}),
and excite a substantial gravity-wave response that may be detectable
even in main-sequence stars (e.g., \citealt{2022A&A...667A..43B}).

Within the radiative zone itself, the latest simulations
(\citealt{2023Sci...379..300P}, \citealt{2023MNRAS.521.5372J}) now
appear to be capturing some aspects of the long-envisioned
``Tayler-Spruit'' dynamo. This differs in some important respects from
the picture originally envisioned by Spruit and debated in
\cite{2007A&A...474..145Z}; for example, in the
\citet{2023Sci...379..300P} simulations only \emph{subcritical} dynamo
action is found, alongside some other dynamo instability. The
saturation amplitude of the field is still very much under debate; see
discussions in \citet{Spruit:1999vt}, \citet{2019MNRAS.485.3661F}, and
\citet{2023MNRAS.521.5372J}.

\subsubsection{Low-mass stars and the transition to full convection}
\label{sec:fullco}

Main-sequence stars less massive than the Sun have deeper convective
envelopes (as a fraction of the total stellar radius). Below a mass of
about 0.35~$M_{\odot}$ stars are -- in standard 1D models --
convective throughout their interiors. This transition occurs at a
spectral type of around M3, so ``M dwarfs'' in general hold special
interest theoretically, as probes of the various roles that rotation
and stratification play in the dynamo. They are also interesting
astronomically: the large majority of stars in our galaxy are M
dwarfs, and they are popular targets in the quest to find and
characterise exoplanets (e.g., \citealt{2018A&A...609A.117T}).

Here, we briefly review attempts to model these stars numerically. In
terms of fundamental fluid dynamics, these stars have many
similarities with pre-main sequence stars (which are also fully
convective, but can have different internal heating profiles and
rotational constraints), so in places our discussion parallels that in
\Sec{sec:ttauri}. In some other respects the flows in these objects
resemble those in giant gaseous planets, so we also draw comparisons
to the extensive literature on planetary dynamos.  Finally, there are
also many similarities with the flow in massive-star cores, which
share the same geometry (i.e., a full sphere of convection) but are
much less strongly stratified than a main-sequence M dwarf; see
\Table{tab:dimpar}.

The first 3D MHD simulations that aimed specifically to model low-mass
fully convective stars were reported in \citet{DSB06}. They used the
{\sc Pencil Code}, solving the fully compressible equations to model a
spherical star (established via volumetric heating and cooling terms)
embedded in a Cartesian grid. These first models included only a
fairly weak density stratification (with $\rho$ at the center of the
star about a factor of three greater than at its photosphere). Later,
\citet{2008ApJ...676.1262B} conducted the first anelastic simulations
(with the {\sc ASH} code) that mimicked low-mass M dwarfs, including a
stronger density stratification (about a factor of 100 across the deep
spherical shell) and more complex flows.  Subsequent work has sampled
much lower diffusivities, more extreme density stratifications, and
varying rotational influences. We note in particular the simulations
of \citet{GDW12} and \citet{YCMGRPW15, 2016ApJ...833L..28Y}, modelling
anelastic dynamos in a deep spherical shell with the {\sc MagIC code};
\citet{2020ApJ...902L...3B}, who considered a full spherical geometry
(i.e., in spherical coordinates but with no singularity at $r=0$)
using the {\sc Dedalus} framework; \citet{2021A&A...651A..66K}, who
considered ``star-in-a-box'' models akin to those of \citet{DSB06} but
in a substantially different parameter regime; and
\citet{2020ApJ...893..107B, 2022ApJ...928...51B}, modelling deep
(anelastic) shells of convection with the {\sc Rayleigh} code.

Although there are many differences between these models, some broad
trends are now reasonably clear, and largely parallel those realised
in other objects and geometries (as discussed elsewhere in this
review). When magnetism is weak or absent, both ``solar'' and
``anti-solar'' differential rotation can be realised, depending upon
the rotational influence (i.e., some version of the Rossby
number). When magnetism is strong, it reduces this differential
rotation and -- if the field gets strong enough -- can essentially
eliminate it, leading to solid-body rotation. The spatial structure of
the field -- e.g., the fraction of the magnetism in axisymmetric
components, or the strength of the dipole or quadrupole moment -- is
intertwined with rotation, shear, and density stratification in a
complex manner.

When rotation is strong and shear is weak, the field tends to develop
a large-scale dipolar component; see discussions in, e.g.,
\citet{GDW12,SPD12,YCMGRPW15}, and an array of related works modeling
planetary dynamos with weak stratifications (e.g., \citealt{CA06},
\citealt{2021GeoJI.224.1890S}). In real stars, presumably any
large-scale field generation is accompanied by vigorous SSD, so the
overall field likely consists of a very wide range of scales. The
dipolar solutions are, at least in some parameter regimes, more
difficult to realise when the density stratification is strong and
when the convective supercriticality is high (e.g., \citealt{GDW12});
but there are now multiple examples of highly-stratified, vigorous
convection that exhibit strong dipoles (e.g., \citealt{YCMGRPW15}),
including some at surprisingly modest rotational constraints (e.g., up
to $\Ro \sim 0.4$ in \citealt{2022MNRAS.517.3392Z}). The question of
what exactly delineates these states from one another is a topic of
very active investigation (see, e.g., discussions in
\citealt{2020PEPI..30706542M}, \citealt{2021GeoJI.226.1897T},
\citealt{2022MNRAS.517.3392Z}).

When shear is also present, a variety of solutions are possible,
including propagating dynamo waves. Examples abound; see, for example,
\citet{2016ApJ...833L..28Y}, \citet{2021A&A...651A..66K}, and
\citet{2022ApJ...928...51B}. Again, it seems reasonably clear that the
rotational influence is the most crucial control parameter, but many
details remain unclear. Other topics of active interest include the
prevalence of non-axisymmetric features in the field (e.g.,
\citealt{2022ApJ...928...51B, 2021A&A...651A..66K}), or modes in which
the bulk of the magnetism is confined to one hemisphere (e.g.,
\citealt{2009PhRvE..80c5302G,GDW12, 2020ApJ...902L...3B}).
Furthermore, \cite{2021A&A...651A..66K} reported that the qualitative
succession of dynamo modes as a function of rotation appears to be the
same in simulations of fully and partially convective stars: when
rotation increases, the predominantly axisymmetric steady and cyclic
solutions at slow rotation give way to non-axisymmetric dynamos at
rapid rotation. In these models a similar succession happens with the
cycles, so that for moderate rotation the dynamo waves typically
propagate in latitude; when rotation is more rapid, the large-scale
magnetic structure drifts in longitude.

\section{Connections to mean-field dynamo theory}\label{sec:conMF}

Finding out which physical processes lead to the observed magnetic
field evolution in 3D convective dynamo simulations is very
challenging. An often-used approach is to interpret the outcome of the
simulations in terms of mean-field dynamo theory. Mean-field theory
provides a well-established theoretical foundation, which can be used
to analyses the complex 3D simulation in a simplified way
\citep[e.g.][]{KR80,BS05}; see also
\cite{2023arXiv230312425B}. Technically this is done by computing
mean-field transport coefficient from 3D simulations and using them in
a corresponding mean-field model. This approach has been successfully
used to pinpoint the cause of magnetic field evolution in many
simulations. In mean-field theory the magnetic and velocity fields are
divided into a mean or averaged part and a fluctuation,
e.g.\ $\BBB=\mBBB + \fluc{\BBB}$. Only the mean fields are explicitly
solved for, whereas the (correlations of) fluctuations are
parameterized in terms of the mean. Azimuthal averages are often used
for solar and stellar dynamos such that the resulting mean field is
axissymmetric. However, such an average is not well suited for rapidly
rotating stars, where non-axisymmetric $m=1, 2$ modes often dominate
\citep{2018A&A...616A.160V}. Applying the mean-field approach to the
induction equation leads to the emergence of an additional term, the
electromotive force
$\mEMF=\overline{\fluc{\uuu}\times\fluc{\BBB}}$. In mean-field dynamo
theory this term is parameterized in terms of the $\mBBB$ and its
gradients, assuming that the mean fields vary slowly in space and time
\citep{KR80},
\begin{equation}
  \EMFi = a_{ij} \mBBj + b_{ijk} \frac{\pd\mBBk}{\pd x_j} + \ldots,
\end{equation}
where the dots represent higher order derivatives that are most often
neglected. The tensors $\aaa$ and $\bbb$ can then be further divided
into symmetric and anti-symmetric parts
\citep[e.g.][]{1980AN....301..101R}, yielding an equivalent
representation
\begin{equation}
\mEMF=\aalpha\bm\cdot\mBBB+\ggamma\times\mBBB 
-\bbeta\bm\cdot(\nab\times\mBBB)
-\ddelta\times(\nab\times\mBBB) 
-\kkappa\bm\cdot(\nab\mBBB)^{(s)},
\label{eq:EMF2}
\end{equation}
where $(\nab\mBBB)^{(s)}$ is the symmetric part of the magnetic field
gradient tensor. The coefficients $\aalpha$ and $\bbeta$ are rank two
tensors, $\ggamma$ and $\ddelta$ are vectors, and $\kkappa$ is a rank
three tensor. These coefficients can be associated with different
turbulent effects important for the magnetic field evolution: the
$\alpha$ effect \citep{1966ZNatA..21..369S} leads to field
amplification via helical flows; $\gamma$ describes the turbulent
pumping, which acts like mean flow
\citep[e.g.][]{1968ZNatA..23.1851R,1975AN....296...49R}; $\beta$
describes turbulent diffusion; and the $\delta$ effect, also known as
the R\"adler effect \citep{1969VeGG...13..131R} or the shear-current
effect \citep[e.g.][]{2003PhRvE..68c6301R}, can lead to dynamo action
in non-helical turbulence in the presence of shear; and finally, the
$\kappa$ effect, whose physical interpretation is currently unclear.

The main challenge is to determine these coefficients from a 3D global
convective dynamo simulation. The ultimate goal is to be able to
reproduce the magnetic field solution with a mean-field model using
the obtained coefficients.

\subsection{Using proxies based on flow and magnetic field properties}

The simplest approach to compare 3D convection simulations with
mean-field theory is to use approximate proxies of turbulent transport
coefficients based on the flow and magnetic field properties.
Assuming isotropy and homogeneity and applying the second order
correlation approximation (SOCA), the turbulent transport tensors
reduce to scalars $\alpha_{\rm K}$ and $\beta$,
\citep[e.g.][]{KR80,2023arXiv230312425B}
\begin{eqnarray}
\alpha_{\rm K} = -{1\over 3} \tau \overline{\fluc{\ooo}\bm\cdot\fluc{\uuu}},\ \ \ \beta  =  {1\over 3} \tau \overline{\uuu^{\prime 2}},
\end{eqnarray}
where $\tau$ is the correlation time of the flow,
$\fluc{\ooo}=\nab\times\fluc{\uuu}$. The expressions of $\alpha_{\rm
  K}$ and $\beta$ are valid in the kinematic regime where the
back-reaction of the magnetic field on the flow is neglected. In a
more general approach, the minimal tau-approximation, this
backreaction is retained and this leads to an additional magnetic
contribution to the $\alpha$ effect \citep{PFL76,2002PhRvL..89z5007B}
\begin{eqnarray}
\alpha_{\rm M}={1\over 3} \tau \mrho^{-1} \overline{\fluc{\JJJ}\bm\cdot\fluc{\BBB}}, \label{eq:alphaM}
\end{eqnarray}
where $\fluc{\JJJ}=\mu_0^{-1}\nab\times\fluc{\BBB}$. $\alpha_{\rm M}$
can be interpreted as a consequence of magnetic helicity conservation
\citep[see, e.g.][]{2023arXiv230312425B}. This approach has been used
in many simulation studies to interpret the magnetic field evolution
\citep{2020LRSP...17....4C}. For example, \cite{WKKB14} used the
$\alpha$ proxy to conclude that the equatorward migration found in
their, and in previous work, is due to an $\alpha\Omega$ Parker dynamo
wave driven by a region of negative radial shear. Similarly,
\cite{DWBG16}, explained the equatorward migration in their
simulations by the inversion of $\alpha$ and positive radial
shear. \cite{GSdGDPKM15,2019ApJ...880....6G} concluded that the
dynamos in their simulations with tachoclines were driven by
$\alpha_{\rm M}$ below the convection zone.  However, it is necessary
to bear in mind the approximate nature of analyses based on proxies:
often the agreement between the simulation and the behavior suggested
by the proxy is qualitative at best and the 3D simulations contain a
rich variety of non-linear interactions that are omitted in such
analyses.

\subsection{Direct measurements of coefficients}

The alternative is to measure the coefficients in \Eq{eq:EMF2}
directly from simulations. There are currently two commonly used
methods for this. First, $\mBBB$ and $\mEMF$ from the 3D dynamo
simulation can be used to fit for the turbulent transport coefficients
in \Eq{eq:EMF2} using, e.g., multidimensional regression method or
singular value decomposition (SVD)
\citep{2002GApFD..96..319B,RCGBS11}. On the other hand, the test field
(TF) method uses a sufficiently large number of linearly independent
\emph{test fields}, that do not back-react on the solution, and
evolves the corresponding $\fluc{\BBB}$ and $\mEMF$ for each. Then it
is possible to unambiguously invert for the coefficients in
\Eq{eq:EMF2} \citep{SRSRC05,SRSRC07}. Both of these methods are, at
best, only as good as the approximate equation \Eq{eq:EMF2}. The
validity of the results needs to be tested be inserting the derived
coefficients back into \Eq{eq:EMF2} and to mean-field models to
determine how faithfully they capture the $\mEMF$ and time evolution
of the mean field in the 3D simulation.

The SVD method has the issue that \Eq{eq:EMF2} is underdetermined:
there are 27 unknown parameters and only three components of $\mBBB$
and $\mEMF$. This is typically overcome by considering the time
dependence of $\mBBB$ and $\mEMF$ leading to an overdetermined system.
Furthermore, if $\mBBB$ does not vary in time the SVD method has
problems to converge. Despite these difficulties this method has been
used to explain the dynamos in several simulations
\citep[e.g.][]{RCGBS11,2013ApJ...777..153A,ABMT15}.
\cite{2016AdSpR..58.1522S} found that the coefficients related the
gradients of $\mBBB$ ($\bbeta$, $\ddelta$, $\kkappa$) are much less
important than $\aalpha$ and $\ggamma$. Furthermore, \cite{SCB13}
could reproduce the mean-field evolution of 3D global simulation using
$\aalpha$ and $\ggamma$ determined with the SVD method, basically
assuming it is an $\alpha^2\Omega$ dynamo. However they had to assume
a higher turbulent diffusivity than what was measured. In follow-up
studies the authors could explain and reproduce a dual dynamo action
\citep{2016ApJ...826..138B} and generate Grand Minima-like events by
including $\alpha$ quenching \citep{SC20}. Recently,
\cite{2022ApJ...935...55S} analysed the simulations of \cite{HRY16}
with the SVD method and found that the turbulent diffusion decreases
with increasing $\ReM$. However, it is unlikely that \Eq{eq:EMF2}, and
hence the SVD method, is valid at high $\ReM$ where a SSD is excited.

\begin{figure*}[t!]
\begin{centering}
 \includegraphics[width=0.9\textwidth]{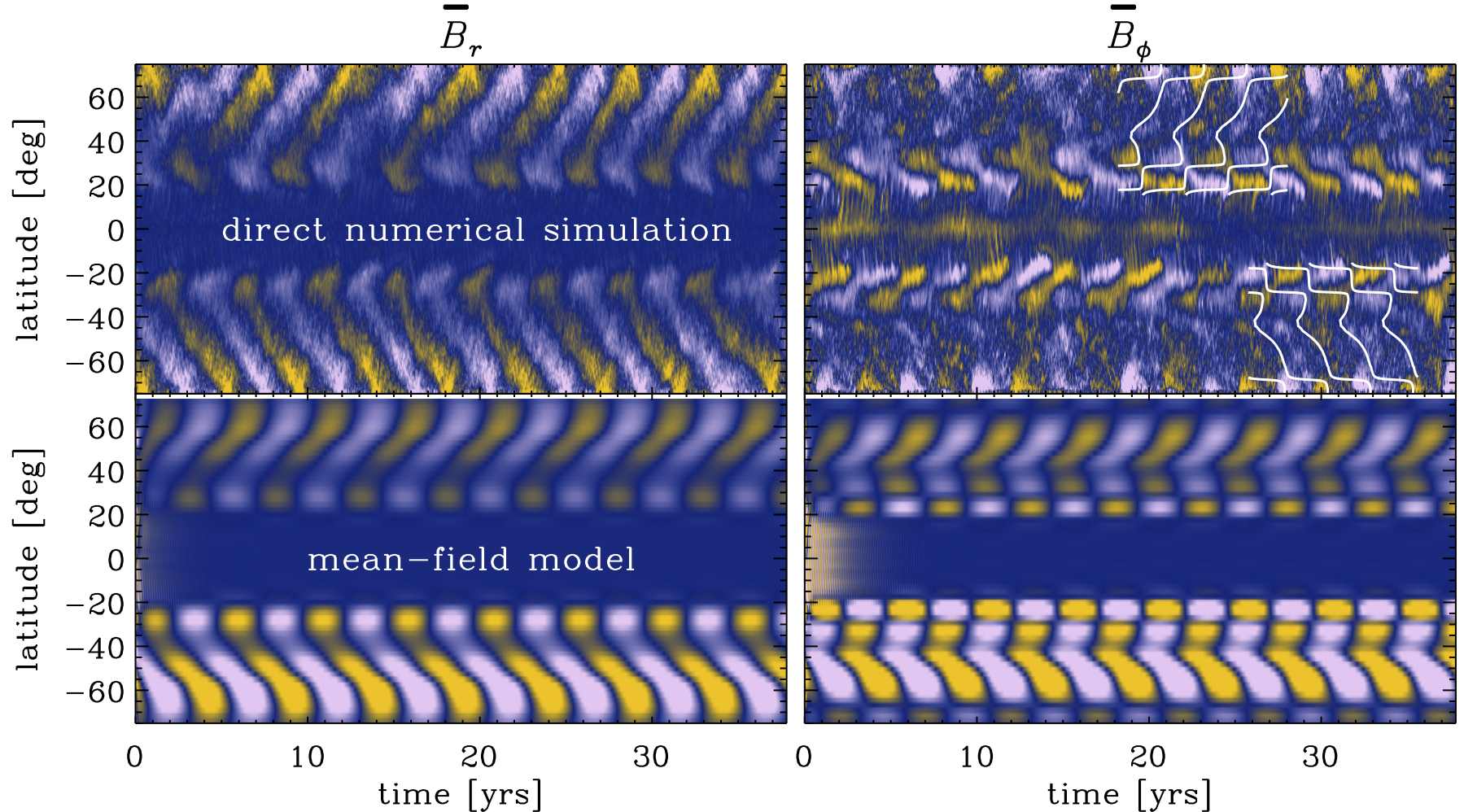}
\caption{Comparison of direct numerical simulations of
  \cite{2018A&A...616A..72W} and \cite{2020A&A...642A..66W} (top) with
  a corresponding mean-field model (bottom) where the turbulent
  transport coefficients have been obtain with the test field method.
  The radial (left) and toroidal (right) mean field is shown as a
  function of latitude and time. The white contours on the top right
  panel indicate the corresponding field of the mean-field
  model. Adapted from \cite{2021ApJ...919L..13W}.}
\label{fig:dns_mf}
\end{centering}
\end{figure*}

In the TF method, a set of 9 linearly independent test fields are used
to uniquely determine the 27 unknowns. First developed for the
geodynamo \citep{SRSRC05,SRSRC07,SPD11,SPD12,Sch11}, it has
subsequently been used for many solar and stellar dynamo simulations
\citep{GKW17,2018A&A...609A..51W,2018A&A...616A..72W,2019ApJ...886...21V,2020A&A...642A..66W}.
An important result is that the turbulent pumping is typically larger
than the meridional circulation in global convective dynamo
simulations, rendering the flux-transport dynamo scenario unlikely in
those cases \citep{2018A&A...609A..51W}. The conclusion of
\cite{WKKB14} that the equatorward migration in these kind of
simulation is explained by a Parker dynamo wave was confirmed with the
$\alpha$ effect from the TF method
\citep{2018A&A...609A..51W,2021ApJ...919L..13W}. This was later used
to explain the cycle period dependence on rotation in 3D dynamo
simulations \citep{2018A&A...616A..72W}. \cite{GKW17} analyzed the
simulations of \cite{KKOBWKP16} and found that the turbulent transport
coefficients -- particularly $\ggamma$ -- vary significantly during
long-term modulation of the cyclic mean magnetic field. In the work of
\cite{2019ApJ...886...21V}, the first cyclic dynamo in the anti-solar
differential rotation regime was explained to be of $\alpha^2\Omega$
type using test-field coefficients. Furthermore,
\cite{2020A&A...642A..66W} studied the transport coefficients as
functions of rotation rate and found that $\alpha\Omega$ dynamos,
appear to be possible in a relative narrow range in $\Ro$. The trace
of $\aalpha$ with $\alpha_{\rm K}$ agree in pattern and amplitude in a
Rossby number range spanning three orders of magnitude
\citep{2020A&A...642A..66W}. As a result of the test-field analysis
\citep{2020A&A...642A..66W} a magnetic influence on $\alpha$ as
described in \Eq{eq:alphaM} could be ruled out in their simulations.
Putting all the turbulent transport coefficients into a mean-field
model, the evolution of the mean magnetic field of the 3D simulation
was reproduced in terms of period and pattern
\citep{2021ApJ...919L..13W}; see \Fig{fig:dns_mf}.  Notably the full
spectrum of coefficients was needed to fully reproduce the field
evolution. This suggests that all of the turbulent mean-field effects
play important roles in this simulation which is a good representation
of current global dynamo simulations.  Furthermore, the authors
concluded that the assumptions of \Eq{eq:EMF2} are reasonably well
justified in the simulations \citep{2021ApJ...919L..13W} given that
the $\EMF$ is also reproduced reasonably satisfactorily
\citep{2018A&A...609A..51W,2019ApJ...886...21V}.

\subsection{Remaining issues}

One of the important issues is that the results of the two methods,
SVD and TF, do not fully agree with each other. The SVD method seems
to produce satisfactory results for EULAG-MHD simulation
\citep{SCB13,2016ApJ...826..138B,SC20}. However,
\cite{2018A&A...609A..51W} showed that the coefficients determined
with the TF and SVD methods give quite different results and that the
SVD coefficients related to the derivative of $\mBBB$ are indeed less
important.  This discrepancy raises questions regarding the overlap of
the validity ranges and the underlying assumptions of both methods.  A
possible reason for the discrepancy is that non-locality (in space
and/or time), which is neglected in \Eq{eq:EMF2}, plays a role: the
SVD uses the actual field whereas in TF only very large-scale
gradients of test fields are retained. On the other hand, the actual
magnetic fields the SVD uses are not necessarily linearly independent,
leading to errors in the inversion.

SVD and TF methods reveal that turbulent transport effects play an
important role in the dynamic and evolution of the large-scale
magnetic field. \cite{2021ApJ...919L..13W} showed that it is necessary
to include practically all of the possible turbulent effects to
reproduce the 3D simulation results in detail. This makes the
corresponding mean-field models quite complex which to a certain
extent defeats the purpose of mean-field modelling where the hope has
been to capture the large-scale behavior of complicated 3D systems by
a much simpler lower-dimensional model. However, the complexity of
dynamos operating in these simulation also hints in the direction that
the Sun and other stars are more complex than simple mean-field models
including the Babcock-Leighton model can describe.

Another issue is related to the appearance of the SSD instability when
global simulations reach more realistic high-$\ReM$ regimes. In this
case \Eq{eq:EMF2} is no longer valid, because a contribution to
$\mEMF$ that is independent of $\mBBB$ is possible, and non-linearity
due to $\fluc{\BBB}$ becomes important.  If small-scale magnetic
fields due to the SSD are dynamically important, neither the SVD nor
TF method will work in their current form. Efforts to generalize the
TF method to incorporate the effect of the SSD have been taken by
\cite{2022ApJ...932....8K}, leading to four flavours of the method in
that regime. Although these flavours should in principle agree, this
is not always the case, especially for high $\ReM$.  Hence, it is of
great importance to extend the SVD and TF methods to more realistic
parameter regimes to incorporate effects such as the SSD.

\section{Conclusions and future prospects}\label{sec:conclusions}

A central challenge in simulating Solar or stellar dynamos is that, as
discussed above, the interiors of real stars are characterised by
extremely low diffusivities (of momentum, heat, and magnetism), and
possess motion and magnetism over an extraordinarily broad range of
spatial and temporal scales.  No simulation, now or in the near
future, can capture all these scales simultaneously.  The hope of many
modelers, though, is that at least \emph{some} aspects of the
dynamics, particularly on the largest scales, may become independent
of the small-scale details at high enough resolution (low enough
diffusivity).  There is considerable debate -- even amongst the
authors of this review! -- about the extent to which present-day
simulations are nearing this diffusion-free, resolution-independent
regime, and reasons for both optimism and pessimism. Here, we briefly
highlight a few of these.

Current global dynamo simulations of stars routinely capture
solar-like differential rotation and cyclic magnetism. Sometimes these
models also reproduce equatorward migrating activity akin to the
Sun. This occurs at a Rossby number regime where differential rotation
is relatively strong. These results seem to be fairly robust
irrespective of the numerical method or other details of the
simulations. Also the theoretical understanding of the physical
mechanisms driving the magnetism has developed significantly in the
recent years with more advanced analysis tools such as the test field
method, full energy transfer and field production (emf) analysis, and
with direct comparisons to mean-field models.

Although our understanding of the physics of convection and resulting
dynamo action has increased, new challenges have also been
encountered. The most intriguing of these is the fact that current
simulations struggle to reproduce solar convection and the resulting
differential rotation at the solar luminosity and rotation rate.
Given that this is a necessary requirement to get the dynamo right it
is not a huge surprise that reproducing the solar dynamo remains
challenging. This ``convective conundrum'' is the modern equivalent to
the ``dynamo dilemma'' of the 1980s. The latter lead to a revival of
old, and the conception of new, ideas about solar and stellar dynamos
and a similar process is at work with respect to solar and stellar
convection at the moment. The various ideas related to solving the
conundrum include entropy rain and deep weakly subadiabatic
convection, the influence of strong small-scale magnetism, and
rotationally constrained deep convection. Research on this topic is
very active and evolving rapidly, and, far from being stumped by the
challenge posed by the convective conundrum, activity in the modelling
of stellar convection and dynamos has instead been invigorated.

A key issue with the Sun is that even though the deep convection zone
is highly likely rotationally dominated with $\Ro\ll1$, there are many
scales in the upper convection zone and near the surface where the
rotational influence on the flow is weak. The Sun is also perhaps
close to the transition to anti-solar differential rotation, which has
some observational support, making it difficult to maintain a
solar-like differential rotation profile if the simulations are not
sufficiently near the correct parameter regime. Identifying the
correct force balance prevailing in the solar convection zone is
therefore key to this problem. With this information, simulations can
be designed such that they follow a path leading to the correct
balances and hopefully to solar-like results. This is a practice
adopted in simulations of the geodynamo and perhaps a similar approach
can be adopted for the Sun.

On the other hand, the issue regarding rotational influence is not as
severe in stars that rotate more rapidly than the Sun and which are
further away from the solar-like to anti-solar differential rotation
transition. There are indications that simulations capture the
characteristics of dynamos, such as non-axisymmetric large-scale
fields, and dipole dominated dynamos in M dwarfs, in such stars more
accurately.  Although this is encouraging, we should also bear in mind
that the observational data from other stars is not as accurate and
detailed as the data we have from the Sun.

A common characteristic of all of the current simulations is the fact
that it is not possible to model the surface layers, where the density
drops vertiginously, accurately enough. This could be one of the
reasons why none of the current simulations form spots that could play
a role in the dynamo process via a Babcock--Leighton type effect, and
their magnetic activity does not become independent of rotation at
sufficiently low $\Ro$. Self-consistent spot formation has not been
reported even in local simulations to say nothing about global
simulations. Therefore capturing spot formation in global simulations
is perhaps as challenging, or even more challenging, than cracking the
convective conundrum.  Nevertheless, recent spot formation studies in
more idealised simulation setups serve as a guide for the design of
future global simulations that aim at achieving this.

All of these developments happen on a background where modelers have
started to realize that the holy grail of stellar dynamo simulations
-- an asymptotic regime where results are independent of resolution or
diffusivity -- remains elusive, and that the computational cost of
adding another data point at a higher resolution is already
prohibitive. This begs the question whether it is feasible for
everyone to try to beat everyone else in this very difficult task or
whether it is better to combine resources for a collaborative effort
where the resources of at least a large part of the field are directed
in producing the ``next generation'' transformative simulations.

\section*{Acknowledgements}

PJK acknowledges the financial support by the Deutsche
Forschungsgemeinschaft Heisenberg programme (grant No.\ KA 4825/4-1).
This project has received funding from the European Research Council
(ERC) under the European Union's Horizon 2020 research and innovation
program (Project UniSDyn, grant agreement no 818665) (JW). This work
was done in collaboration with the COFFIES DRIVE Science
Center. A.S.B. acknowledges support by European research council for
ERC grant Stars2 (grant no.\ 207430) and Whole Sun (grant
no.\ 810218), by CNRS/INSU through the Sun-Earth French national
program, by French space agency CNES with Solar Orbiter grant and
local fundings by Observatory OSUPS, Universities of Paris-Saclay and
Paris-Cit\'e. GG acknowledges support from NASA grants NNX14AB70G,
80NSSC20K0602, and 80NSSC20K1320.

%\bibliography{bib}% common bib file

\input paper.bbl
\end{document}

%% file: macros.tex
\newcommand{\mrho}{\overline{\rho}}

\newcommand{\aaa}{{\bm a}}

\newcommand{\bbb}{{\bm b}}

\newcommand{\uuu}{{\bm u}}

\newcommand{\ooo}{{\bm \omega}}

\newcommand{\BBB}{{\bm B}}
\newcommand{\mBBB}{\overline{\bm B}}

\newcommand{\FFF}{{\bm F}}

\newcommand{\JJJ}{{\bm J}}

\newcommand{\UUU}{{\bm U}}

\newcommand{\gggg}{{\bm g}}

\newcommand{\mBBj}{\overline{B}_j}
\newcommand{\mBBk}{\overline{B}_k}

\newcommand{\mEMF}{\overline{\bm{\mathcal{E}}}}

\newcommand{\EMFi}{\overline{\mathcal{E}}_i}
\newcommand{\Eq}[1]{Eq.~(\ref{#1})}

\newcommand{\EN}{\end{equation}}
\newcommand{\EQA}{\begin{eqnarray}}
\newcommand{\ENA}{\end{eqnarray}}

\newcommand{\pd}{\partial}

\newcommand{\mean}[1]{\overline{#1}}

\newcommand{\cP}{c_{\rm P}}
\newcommand{\cV}{c_{\rm V}}
\newcommand{\cs}{c_{\rm s}}

\newcommand{\urms}{u_{\rm rms}}
\newcommand{\uconv}{u_{\rm conv}}

\newcommand{\orms}{\omega_{\rm rms}}

\newcommand{\Pcyc}{P_{\rm cyc}}
\newcommand{\Prot}{P_{\rm rot}}

\newcommand{\Ma}{{\rm Ma}}

\newcommand{\sigmaSB}{\sigma_{\rm SB}}

\newcommand{\chiSGS}{\chi_{\rm SGS}}

\newcommand{\Co}{{\rm Co}}
\newcommand{\Cow}{{\rm Co}_\omega}

\newcommand{\HP}{H_{\rm P}}

\newcommand{\Pe}{{\rm Pe}}
\newcommand{\Pra}{{\rm Pr}}

\newcommand{\PraSGS}{{\rm Pr}_{\rm SGS}}

\newcommand{\PrM}{{\rm Pr}_{\rm M}}

\newcommand{\Ra}{{\rm Ra}}

\newcommand{\Rey}{{\rm Re}}

\newcommand{\Roc}{{\rm Ro}_{\rm c}}

\newcommand{\ReM}{{\rm Re}_{\rm M}}
\newcommand{\Ro}{{\rm Ro}}

\newcommand{\Ta}{{\rm Ta}}
\newcommand{\Ek}{{\rm Ek}}
\newcommand{\taun}{\tau_{\rm n}}

\newcommand{\tauKH}{\tau_{\rm KH}}

\newcommand{\taucyc}{\tau_{\rm cyc}}

\newcommand{\tauconv}{\tau_{\rm conv}}
\newcommand{\nab}{\mbox{\boldmath $\nabla$} {}}

{}

\newcommand{\Rgas}{{\cal R}}
\newcommand{\EMF}{\mbox{\boldmath ${\cal E}$} {}}
\def\onethird{{\textstyle{1\over3}}}

\def\onehalf{{\textstyle{1\over2}}}

\newcommand{\Fig}[1]{Figure~\ref{#1}} %for ApJ
\newcommand{\Sec}[1]{Section~\ref{#1}} %for ApJ

\newcommand{\Table}[1]{Table~\ref{#1}}

\newcommand{\fluc}[1]{#1^\prime}
\newcommand{\ddelta}{\mbox{\boldmath $\delta$} {}}
\newcommand{\ggamma}{\mbox{\boldmath $\gamma$} {}}
\newcommand{\aalpha}{\mbox{\boldmath $\alpha$} {}}
\newcommand{\bbeta}{\mbox{\boldmath $\beta$} {}}
\newcommand{\kkappa}{\mbox{\boldmath $\kappa$} {}}

%% file: journals.tex
\def\aj{AJ}                     % Astronomical Journal
\def\apj{ApJ}                   % Astrophysical Journal
\def\apjl{ApJ}                  % Astrophysical Journal, Letters
\def\apjs{ApJS}                 % Astrophysical Journal, Supplement
\def\aap{A\&A}                  % Astronomy and Astrophysics
\def\aapr{A\&A~Rev.}            % Astronomy and Astrophysics Reviews
\def\icarus{Icarus}             % Icarus

\def\mnras{MNRAS}               % Monthly Notices of the RAS
\def\pre{Phys.~Rev.~E}          % Physical Review E
\def\solphys{Sol.~Phys.}        % Solar Physics
\def\ssr{Space~Sci.~Rev.}       % Space Science Reviews
\def\zap{ZAp}                   % Zeitschrift fuer Astrophysik
\def\nat{Nature}                % Nature
\def\memsai{Mem.~Soc.~Astron.~Italiana}
\def\physrep{Phys.~Rep.}        % Physics Reports